\documentclass[fleqn,usenatbib]{mnras}
\usepackage{newtxtext,newtxmath}

\usepackage[T1]{fontenc}
\usepackage{ae,aecompl}

\usepackage{graphicx}	
\usepackage{amsmath}	
\usepackage{amssymb}	
\usepackage[range-units=single,separate-uncertainty,detect-mode]{siunitx}
\usepackage{booktabs}

\usepackage{comment}
\usepackage{blindtext}

\title[P-REx: The Piston Reconstruction Experiment]{P-REx: The Piston Reconstruction Experiment for Infrared Interferometry}

\author[F. Widmann et al.]{
Felix Widmann,$^{1}$\thanks{The presented research was done at MPIA in Heidelberg. However, first author F.W. moved on during paper submission to the Max-Planck-Institute for extraterrestrial Physics, Giessenbachstr., in D-85748 Garching, Germany, and can be contacted by e-mail: fwidmann@mpe.mpg.de}
J\"{o}rg-Uwe Pott,$^{1}$
Sergio Velasco$^{2,3}$ \\
$^{1}$ Max-Planck-Institute for Astronomy (MPIA), K\"{o}nigstuhl 17, D-69117 Heidelberg, Germany\\
$^{2}$ Instituto de Astrof\'{i}sica de Canarias, c/V\'{i}a L\'{a}ctea s/n, La Laguna E-38205, Spain\\
$^{3}$ Departamento de Astrof\'{i}sica, Universidad de La Laguna, La Laguna E-38200, Spain
}

\date{Accepted XXX. Received YYY; in original form ZZZ}

\pubyear{2017}

\begin{document}
\label{firstpage}
\pagerange{\pageref{firstpage}--\pageref{lastpage}}
\maketitle

\begin{abstract}
For sensitive infrared interferometry, it is crucial to control the differential piston evolution between the used telescopes. This is  classically done by the use of a fringe tracker. In this work, we develop a new method to reconstruct the temporal piston variation from the atmosphere, by using real-time data from adaptive optics wavefront sensing: the Piston Reconstruction Experiment (P-REx). In order to understand the principle performance of the system in a realistic multilayer atmosphere it is first extensively tested in simulations. The gained insights  are then used to apply P-REx to real data, in order to demonstrate the benefit of using P-REx as an auxiliary system in a real interferometer. All tests show positive results, which encourages further research and eventually a real implementation. Especially the tests on on-sky data showed that the atmosphere is, under decent observing conditions, sufficiently well structured and stable, in order to apply P-REx. It was possible to conveniently reconstruct the piston evolution in two-thirds of the datasets from good observing conditions (r$_0$ $\sim$ \SI{30}{\centi\meter}). The main conclusion is that applying the piston reconstruction in a real system would reduce the piston variation from around \SI{10}{\micro\meter} down to \SIrange{1}{2}{\micro\meter} over timescales of up to two seconds. This suggests an application for mid-infrared interferometry, for example for MATISSE at the VLTI or the LBTI. P-REx therefore provides the possibility to improve interferometric measurements without the need for more complex AO systems than already in regular use at 8m-class telescopes. 
\end{abstract}

\begin{keywords}
instrumentation: interferometers -- instrumentation: adaptive optics -- infrared: general
\end{keywords}


\section{Introduction}
One of the main complications for ground-based optical and near-infrared interferometry is the random and fast changing piston drift over the individual telescopes, introduced by the atmospheric turbulence. The challenge of fast changing piston values is usually approached by the use of fringe trackers, which measure the movement of the interferometric fringes \citep[see e.g.][]{LeBouquin2008}. Since the atmospheric piston cannot be measured directly, the use of an interferometric fringe tracker is unavoidable. This fundamental need of fringe tracking limits the sensitivity and sky coverage of direct interferometry in the visible and infrared wavelength regime. For modern fringe trackers natural guide stars with magnitudes brighter than 10 in the H-Band are required \citep[see e.g.][]{LeBouquin2008,Choquet2010,Choquet2014}. This restriction is due to the short coherence time of the atmosphere. To overcome this problems, we present an algorithm which can extend the integration time over the atmospheric coherence time. This is particularly timely, since telescope vibrations as the other dominant contribution to piston noise in interferometric instruments are  more and more under control. This is achieved by advanced control algorithms and accelerometer-based vibration measurement systems \citep{Bonnet2006,Choquet2012,Boehm2017}.

With this project, we develop and test an auxiliary method, which uses the real-time data from adaptive optics (AO) wavefront sensing, in order to reconstruct the atmospheric piston drift over a single telescope: the Piston Reconstruction Experiment (P-REx). The core of P-REx is to get the piston drift from temporal AO wavefront information by deriving the dominant wind speed and direction. Combining the wind measurement and atmospheric tip-tilt information, under the assumption of the frozen flow hypothesis \citep{Taylor1938}, gives the piston drift over a single telescope. By doing this measurement at each individual aperture of an interferometer, the difference in piston evolution can be determined for each baseline. With this information, the fringes can be kept stable over short timescales and the integration time of the fringe tracker can be extended. This decreases the problems of fringe trackers, such as high RMS values towards fainter targets \citep{Sahlmann2009} and flux dropouts \citep{LeBouquin2008,Choquet2010}, and effectively increases the coherence time in comparison to the currently implemented approach of direct fringe tracking. An increased coherence time ultimately improves the sensitivity of the interferometer. Furthermore, it could increase the magnitude limit of fringe tracking guide stars, which would open the possibility of optical interferometry for an increasing number of science cases. The scientific goal is therefore to increase the sensitivity of an AO supported infrared interferometer, in order to observe larger, statistically relevant samples of rare objects, like massive young stars and active galactic nuclei. An additional goal is to reach new target classes like brown dwarfs and microquasars, currently out of reach for infrared interferometry. A key advantage of the proposed method is the fact that no additional hardware is needed, if the interferometer is  already equipped with a piston-neutral AO system, fast delay lines, and a fringe tracking system, which is the case for large aperture interferometers like the very large telescope interferometer (VLTI) and the large binocular telescope (LBT). 

In a previous work by \cite{Pott2016}, it has already been demonstrated in idealized simulations that the concept works for multilayer turbulence with uncorrelated wind speeds and turbulence between the layers, due to the typical dominance of the ground layer turbulence. These first simulations showed that P-REx can retrieve the effective wind speed and direction of the atmosphere precisely. In this paper, we expand the simulation to a more realistic end-to-end model to study systematically the requirements to an AO system for the use of P-REx. Then, we conclude our study with first tests on real on-sky data. The article is organized as follows: After an introduction to the details of the piston reconstruction experiment concepts in \autoref{sec:prex}, the method is then intensively tested in realistic end-to-end simulations in \autoref{sec:simulation} and finally applied to on-sky data from the LBT in \autoref{sec:flao}.

\section{The Piston Reconstruction}\label{sec:prex}
P-REx is not able to measure the absolute piston values, it predicts the piston drift with time over each of the telescopes of the interferometer, and thereby the fringe position drift with time. Thus, P-REx is an auxiliary method to help keeping the interferometer coherent, it cannot fully replace a direct fringe measurement, but it shall increase the time scale over which the direct fringe measurement, e.g. by a fringe tracker, needs to be done. By calculating the difference between the two piston drift predictions from each telescopes, we get the differential piston movement between the two telescopes. This differential piston has then to be compensated by the existing delay lines in order to lock the fringe position in the beam combining instrument. P-REx has to be implemented as feed-forward controller in the delay line position controller. The correction has to be done as fast as possible, at least with a frequency of \SI{100}{\hertz}, in order to keep the fringe position locked and not be limited by boiling effects, as will be discussed later. As the piston reconstruction is not error free, an error will add up during the time P-REx is used to stabilize the fringe position. To reduce this error a (slower) fringe tracker has to be used, which will also correct for drifts and calibration errors of the delay line system. The idea is therefore to run the fringe tracker over longer timescales than currently used \citep[$\sim$\SI{100}{\hertz}, see e.g.][]{Sahlmann2009}. A frequency of the order of a few Hertz should be enough to remove the P-REx fringe residual error. With this system, the combination of the usual fringe tracker at low frequency and P-REx at high frequency, the fringes can be stabilized.

The actual P-REx measurement is done for each telescope individually, e.g. by the AO real-time computer (RTC) to avoid transfer large amounts of data in real-time. Since the core of the method is the piston drift prediction at the individual telescopes, the following chapters focus on the piston drift reconstruction at a single telescope only. The differential piston evolution for one baseline is then just the difference of two measurements from the individual telescopes.

\subsection{The Method}
Originally there were two ideas for the reconstruction of the piston drift. The first one uses a reconstructed wavefront in an overlapping area of two frames in order to measure the piston difference of two measurements. The second idea uses only the tip and tilt measurement and the dominant wind vector. As the tip-tilt method has no need for reconstruction the actual wavefront, it has proven to be more accurate and also faster than the first method. Due to these advantages we only describe the tip and tilt method now. For more information on the first method see \citet{Pott2016}.

The concept of piston drift reconstruction is based on the atmospheric data acquired by the wavefront sensor (WFS). In an actual AO system, a feedback control loop is applied. Therefore, one does not get the full phase information from the WFS, but only the residual phase. For the piston reconstruction the full wavefront information is required, which can be achieved by a pseudo open-loop (POL), using the shape of the deformable mirror (DM) and the WFS data. The POL data is calculated with the following formula \citep[e.g.][]{Guesalaga2014}:
\begin{equation}
	S_i^{pol} = S_i^{res} + IM \cdot V_{i-k}
\label{equ:pol}
\end{equation}
In this formula $S_i^{pol}$ is the POL WFS data at a discrete time i. $S^{res}$ is the WFS measurement and $V$ the voltages applied to the DM, which is converted into WFS slope units with the interaction matrix (IM). k is the number of frames by which the application of the voltages to the DM is delayed (usually 1 or 2). These POL data in a 2D representation are used to measure the dominant wind vector. The idea of the piston reconstruction is then that the piston drift is simply the product of the wind velocity and the tip and tilt of the atmosphere:
\begin{align}
\begin{split}
	\Delta P & = \vec{TT}\cdot\Delta s = \vec{TT}\cdot \vec{v}_{wind}\cdot\Delta t \\
	& = \left[\mbox{tip}\cdot v_x + \mbox{tilt}\cdot v_y\right]\cdot \Delta t
\end{split}
\label{equ:prex}
\end{align}
where $\Delta$s is the displacement vector of the wavefront , $\vec{v}$ is the wind vector, and $\Delta t$ is the time over which the piston variation is measured. The tip is the first derivation of the wavefront in x direction and the tilt in y direction. All parameters are again taken from the POL wavefronts, as defines in \autoref{equ:pol}, in order to get the full atmospheric information. 

Under the assumption that the atmosphere is temporally stable and just driven over the telescope by the wind, the piston drift is just the spatial displacement times the tip-tilt. \autoref{equ:prex} can then be used, as the tip and tilt are the derivatives of the spatial piston distribution. However, the calculation is only an approximation. It assumes that the tip and tilt contain the whole phase information. The idea behind this is that a local piston drift is the derivative of the phase, which is given by the tip and the tilt. Another way of looking at the dominant role of tip and tilt is by looking at the power spectrum of atmospheric turbulences \citep{Hardy}. There, the first modes, namely tip and tilt, are the most dominant ones. Therefore the tip-tilt approximation seems to be a valid assumption. Another requirement for P-REx is that the higher order DM modes applied are piston free, which usually is the case for an interferometer \citep[see e.g.][]{Verinaud2001}.

\subsection{Taylors Frozen Flow Hypothesis}
An important assumption for the piston drift reconstruction is the so called Taylor's Frozen Flow Hypothesis (TFFH) first introduced by \citet{Taylor1938}. This hypothesis states that the complete atmosphere can be described as a composition of several different layers. Each of these layers of atmosphere can be described by the Kolmogorov turbulence model \citep{Hardy}, stays spatially stable in time, and is only moved by the wind velocity. The total measured turbulence is the superposition of all these layers. 

There have been different studies in order to verify the frozen flow theory and identify the amount of random fluctuations in the atmosphere. The random and chaotic behavior is called boiling and is the second important aspect in the temporal evolution of the atmosphere \citep{StJacques1998}. The most common approach to test the importance of frozen flow is to use cross-correlation and deconvolution techniques on the data from one or several WFS. \citet{Schoeck2000} came to the result that TFFH accounts for \SI{80}{\percent} of the temporal development for about \SI{20}{\milli\second}. \citet{Guesalaga2014} found similar timescales, adding that the decay rate of the frozen flow correlation increases linearly with increasing wind velocities. Another study was done by \citet{Stjacques2000} with focus on the coherence time of the atmosphere. They find that the coherence peak decreases with time and reaches values as low as 30 \% of its initial value after  \SI{100}{\milli\second}, similar to \citet{Schoeck2000}. A different approach was taken by \citet{Gendron1996,Poyneer2009,Cortes2013} by using a Fourier analysis of the WFS slope measurements. In general their results agree with the previous studies, but \citet{Poyneer2009} also investigated the stability of the velocity vector for each layer. They found much longer timescales, of several minutes or even hours, for the deconvolution peak of a single layer. This shows that TFFH is valid over long timescales for individual layers, but only for short timescales for the whole atmosphere. This agrees with findings by \citet{Avila2006} that the wind profiles stay comparably stable over whole nights. 

In conclusion, basically all performed studies came to the result that TFFH is valid over small timescales \citep{Bharmal2014}. From these studies, we conclude that TFFH is reasonably good as a first approach on timescales below  \SI{20}{\milli\second}. As the piston reconstruction is expected to work in frequencies in the range of \SI{100}{\hertz}, boiling should not be a major limitation.

\subsection{Advantages \& Complications}
The main advantage of the proposed piston reconstruction is that it uses solely the data from the AO system and is therefore more photon-efficient than a fringe tracker. The AO system sits directly at the telescope and does not suffer from photon losses such as the fringe tracker (due to less number of reflections). Another important point is that all the required hardware, such as the AO system and delay lines, are already available at the relevant interferometers. An implementation of the system can therefore be done relatively easy.

There are also some effects that could possibly limit the usability of the system. First of all, there is the problem of atmospheric effects such as boiling and multilayer movements. The consequences of these effects are discussed in this work with on-sky data in \autoref{sec:flao}. Another problem could be additional piston changes that are not detected by the system. These changes could occur for example due to vibrations in the light path between the telescopes. Such effects especially play a role for long baselines, where the piston reconstruction could otherwise be very helpful. An approach to correct such effects can be an additional system, such as the laser metrology system used for GRAVITY \citep{Lippa2016} or accelerometers \citep{Boehm2014}. This article concentrates on understanding the usability of P-REx by analyzing the atmospheric effects and the general performance of a piston reconstruction over a single aperture. In the near future, further work will explain how additional effects from the interferometer can be compensated.

\section{Simulations}\label{sec:simulation}
Throughout this section we use simulated adaptive optics data to test and validate the proposed techniques. These data were produced using the YAO, an end-to-end Monte Carlo simulation software developed by Francois Rigaut \citep{Rigaut2013}. For this work we used the version 5.7.0 with own additions, in order to get the required data from the simulation. YAO is a powerful simulation tool which allows to simulate a wide variety of different AO systems with realistic noise and error contributions. It is fast, widely used, and refined for the past 15 years, making it comparatively bug free and up-to-date with current requirements for AO systems. 

In order not to be limited by dimensioning errors of the AO system, such as a low sampling of the atmosphere or a wind vector estimation precision limited by the available spatial sampling, we start the first simulations with a high sampled AO system. Likewise we started by using a single atmospheric layer with good seeing conditions. For the parameter of the basic simulations see \autoref{tab:aosystem}.

\begin{table}
	\centering
   	\caption[Properties of the AO system for general tests]{Properties of the AO system for general tests.}
	\label{tab:aosystem}
	\begin{tabular}{ll}
		\toprule
		Keyword & Value \\
		\cmidrule(lr{5pt}){1-2}
		Telescope diameter & \SI{8}{\metre}\\
		D/r$_0$ & 40\\
		Wind velocity & \SI{20}{\metre\per\second} \\
		Wavelength & \SI{650}{\nano\metre} \\
		Guide star luminosity & \SI{-5}{mag} \\
		WFS type& Shack Hartmann \\
		Number of WFS lenslets & 20 x 20 \\
		Number of DM actuators & 20 x 20 \\
		Frequency & \SI{500}{\hertz}\\
		\bottomrule
	\end{tabular}
\end{table}

\subsection{Wind Measurement}\label{sec:wind}
As known from signal processing, the displacement between two signals can be determined by calculating the cross-correlation between them \citep{Jaehne2005}. This is also the usual approach for finding the displacement between two measured wavefronts \citep[see e.g.][]{Schoeck2000}. For this reason we use a normalized cross-correlation to measure the shift between two images. The usual normalized cross-correlation between an image f and a template t is defined by:
\begin{align}
\begin{split}
	T_{f,t}(\Delta x,\Delta y) = & \frac{1}{N(\Delta x,\Delta y)} \\
	& \cdot \sum_{x,y} \frac{\left(f(x,y)-\bar{f}\right)\cdot \left(t(x+\Delta x,y + \Delta y)-\bar{t}\right)}{\sigma_f\cdot\sigma_t}
\label{equ:nxcorr}
\end{split}
\end{align}
where $\bar{f}$ and $\bar{t}$ are the averages over the whole images and $\sigma$ is the corresponding standard deviation. N is the so-called overlapping factor, which is equivalent to the number of overlapping pixels for each individual point of the cross-correlation \citep{Schoeck1998}.  This factor compensates the effect that a different number of overlapping pixels is used for each position ($\Delta x,\Delta y$) in the cross-correlation. N can be calculated by doing the auto-correlation of the pupil image. Other sources have introduced a more sophisticated overlap factor \citep[see e.g.][]{Guesalaga2014}, but this is not necessary here, as we are not dealing with additional effects which can occur in real atmospheric images, such as shadows from mirror mounts or traces of laser guide stars. The approach to use a normalized cross-correlation in order to calculate an atmospheric shift is commonly used \citep{Schoeck2000,Wilson2002,Guesalaga2014}.

As the cross-correlation calculates the sum of two two-dimensional arrays, it is comparably slow. There are several approaches to decrease the computational effort of the normalized cross-correlation \citep[see e.g.][]{Lewis1995b,Briechle2001}. The easiest one is to use the relation that the cross-correlation is equal to the inverse Fourier transformation of the cross spectrum \citep{Jenkins1968,Scargle1989,Lewis1995}. Therefore, the normalized cross-correlation can be redefined as:
\begin{align}
	\begin{split}
	T_{f,t}(\Delta x,\Delta y) = & \frac{1}{N(\Delta x,\Delta y)\sigma_f\sigma_t} \\
	& \cdot \mathcal{F}^{-1}\left[\mathcal{F}\left(f(x,y)-\bar{f}\right)\cdot \mathcal{F}^*\left(t(x,y)-\bar{t}\right)\right]
	\label{equ:nxcorrfft}
	\end{split}
\end{align}
where $\mathcal{F}$ is the Fourier transformation and the rest of the notation is identical to \autoref{equ:nxcorr}. N is then as before calculated from the pupil image:
\begin{equation}
	N(\Delta x,\Delta y) = \mathcal{F}^{-1}\left[\left|\mathcal{F}\left(\mbox{pupil}(x,y)\right)\right|^2\right]
\end{equation}
with $\mbox{pupil}(x,y)$ equals 1 inside the pupil and 0 otherwise \citep[see e.g.][]{StJacques1998}.

\subsection{Wind Vector from WFS Measurement}
\begin{figure}
	\includegraphics[width=0.97\columnwidth]{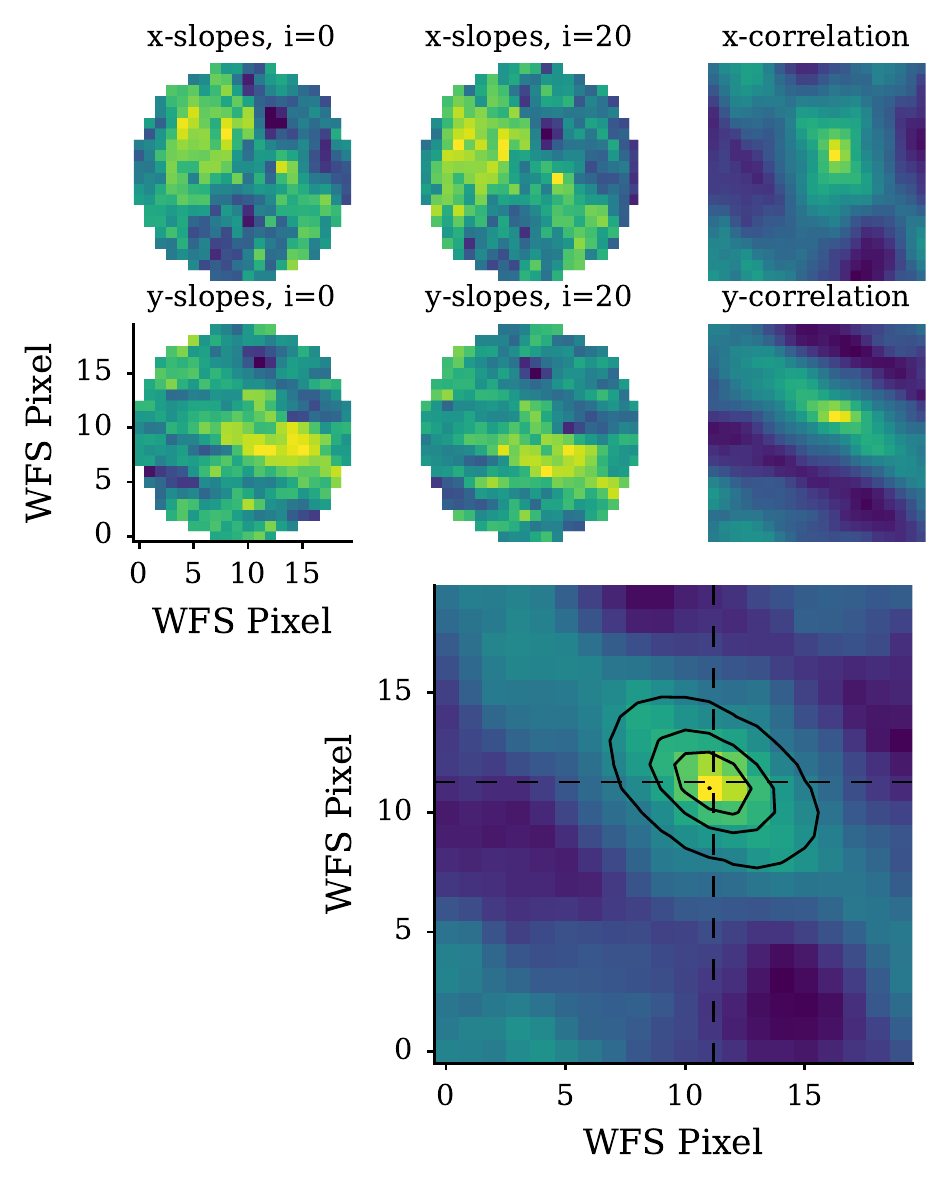}
	\caption{Wind vector detection from SH-WFS slopes. Left two columns: Input slopes of the initial image (i=0), and after 20 time steps (i=20). Third column: Normalized cross-correlations from x- and y-slopes. Large image: Average of the two cross-correlations with a 2D Gaussian fit. The determined wind velocity is 9.8 m/s at 44$^\circ$, with a theoretical value of 10 m/s at 45$^\circ$.}
	\label{fig:nxcorrsh}
\end{figure}
In order to avoid the need of a reconstructed wavefront, the wind detection has to work directly on the WFS data. The simulations are done with slope data from a Shack-Hartmann WFS (SH-WFS). This can be directly adapted to a pyramid WFS, as both types measure wavefront slopes. Another possible WFS is the curvature sensor, used in the MACAO system at the VLTI \citep{Arsenault2003}. Our simulations showed that the wind detection also works with a curvature sensor. However, for the sake of simplicity all the tests in this paper are done with slope data.

In principle, the normalized cross-correlation can be directly applied to the 2D representation of the WFS data to get the wind velocity. The approach is the same as for the cross-correlation with the phase screens, except that the correlation is used individually for the x and y slopes and the average of the two results has to be calculated. However, before doing the cross-correlation, some pre-processing of the data is necessary. As the wind velocity is contained in each individual cross-correlation with a fixed time difference the average over several cross-correlations reduces the noise and the final measurement error. For the simulations we usually used datasets of \SI{10}{\milli\second}, as this is the time scale where one expects TFFH to dominate the turbulence evolution. However, it is also  possible to take longer datasets for the wind measurements and only apply the piston reconstruction to shorter times, as the wind is supposed to be stable over longer time scales \citep{Avila2006,Poyneer2009}. This is done with a moving average and can further decrease the noise in the cross-correlation, which will be especially important for on-sky data.

For each individual slope measurement, we then subtracted the mean value of the x and y slopes, which is the tip and the tilt of the measurement. The tip and tilt have the most power of all modes but do not contain information on the wind vector as they are only a constant factor in the slope measurement. Therefore, the cross-correlation works better on higher modes and the result increases for tip and tilt reduced data. Furthermore, we subtracted the time average of each individual slope measurement over the whole dataset. This is necessary, as static features in the telescope would lead to a permanent peak at the zero point of the cross-correlation \citep{Schoeck1998,Stjacques2000}. Such static features could be for example a bias on individual pixels from a slightly deformed mirror. After these pre-processing steps the normalized cross-correlation as given in \autoref{equ:nxcorrfft} can be applied (for an example see \autoref{fig:nxcorrsh}). 

The sampling of the WFS data is usually rather low, in order to optimize the WFS sensitivity. It is therefore crucial to detect shifts in the cross-correlation which are smaller than one WFS sub-aperture, which corresponds to one pixel in the slope data. The first approach to get the exact shift is to fit a two-dimensional Gaussian to the data (\autoref{fig:nxcorrsh}). This approach works very good and is pretty solid for different kinds of datasets. The downside of such a fit is, that the calculation is comparably slow. \citet{Roopashree2013} did an analysis on the peak detection, with special focus on the extraction of the wind speed from the slopes of a SH-WFS. They came to the conclusion that a ``3-point Parabolic Interpolant'' is the best solution to get the exact shift. In our tests this method tends to overestimate the wind speed, which is also mentioned by \citet{Roopashree2013}. Therefore we ultimately use their second method, the ``3-point Gaussian Interpolant'', as it gives good results in our test and is significantly faster than a complete two-dimensional Gaussian fit. The ``3-point Gaussian Interpolant'' fits a one-dimensional gauss to three pixels, which are the pixel with the maximum value and its two neighbors. By doing this for x and y direction, one can determine the exact peak position.

\subsection{Error estimation}
There are several possible error sources in the WFS data; there is an uncertainty in the slope measurement which depends on the used WFS, as the pixel spacing of the sub-apertures (individual elements of the WFS) and the centroid detection of the spots can vary. Additionally, there are also noise sources from the detector, such as the read-out noise.  All these noise sources are included in the YAO simulation. However, these are most likely not the dominant error of the wind measurement \citep{Schoeck2000}. There are more prominent error sources due to the given data and the cross-correlation. One of them is the very low spatial sampling of the WFS data \citep{Schoeck1998}, which leads to a high pixelation of the data (see \autoref{fig:nxcorrsh}).  This directly leads to an error in the determination of the wind vector, as the peak  position in the cross-correlation can only be detected with an accuracy of a certain fraction of one pixel.  For an individual cross-correlation as shown in \autoref{fig:nxcorrsh}, the uncertainty of the peak position measurement is around a tenth of a pixel. The fitted Gaussian has a FWHM usually slightly larger than one pixel.

By using a larger dataset to calculate the wind vector for 100 individual measurements at stable wind velocity with \SI{20}{\meter\per\second}, we further determined the uncertainty of the algorithm. The results were comparably stable with a standard deviation of 0.06 pixel. This leads to a relative error in the peak position and therefore also in the wind velocity of 10\%. However, this is only the uncertainty of the measurement for a single layer of the atmosphere. If all the involved atmospheric effects such as boiling and a multilayer atmosphere are taken into account, the errors in the measurement will increase. Especially the strength of boiling is very hard to predict and it is therefore difficult to make an appropriate prediction for its error contribution. As mentioned earlier, we circumvent this by estimating the limiting effect of boiling by applying P-REx to real on-sky WFS data in the last part of this work. As a general rule, we can assume that if the SNR per WFS sub-aperture is good enough to keep the AO loop closed, it will be good enough to run P-REx, which effectively averages over the sub-apertures. 

\subsection{First Tests}
For the first complete tests we use the AO system specified in \autoref{tab:aosystem}. The necessary data, namely the WFS slopes, DM voltages and interaction matrix are directly taken from the output of YAO. In order to have a theoretical comparison, the phase screen which YAO sends to the WFS, are also saved. These screens are used to get the theoretical piston and the tip and tilt values, to compare with the results of the P-REx algorithm. 

\begin{figure}
	\includegraphics[width=0.97\columnwidth]{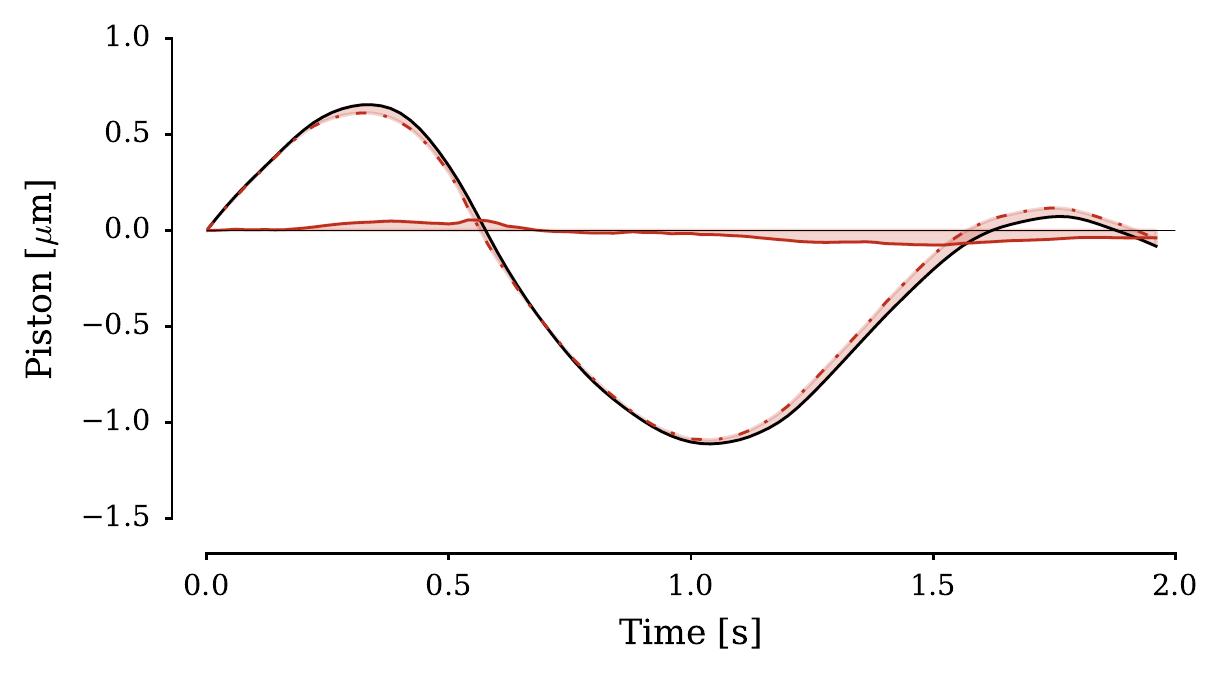}
	\caption{Reconstructed piston from the P-REx algorithm. The data is taken from the simulation of the AO system specified in \autoref{tab:aosystem}. The theoretical value is plotted in black, the red dotted-dashed line shows the reconstructed piston. The residual value is shown as a red solid line and a shaded area.}
	\label{fig:recpiston}
\end{figure}

At this point it is necessary to quantify the performance of the piston reconstruction. In order to do so, we extract the actual piston value from the used phase screens and subtract the initial value, as the algorithm only detect the shift in piston and not the actual piston values. The result from the P-REx algorithm is not the actual piston value, but the differential piston between the two used frames. This differential piston is then used to reconstruct the actual piston, by simply adding up all the differential piston values:
\begin{equation}
	P_{\text{P-REx}}(T) = \sum_{i=0}^{T} dP_{\text{P-REx}}(t=i)
	\label{equ:sum}
\end{equation}
This reconstructed piston can directly be compared with the theoretical piston. An example for the comparison of the theoretical and reconstructed piston is shown in \autoref{fig:recpiston}. 

For the further discussion it is helpful to set up a value in order to quantify the quality of the result from the piston reconstruction. To do so, we calculate the root-mean-square error (RMSE) between the reconstruction and the theoretical value. This is somehow an arbitrary definition for the error of the measurement, as the piston values from the reconstruction are calculated by the summation of the differential piston. This means that every error of the reconstruction also gets summed up, which will ultimately lead to an increasing error. The RMSE therefore increases with time. In order to still get uniform results, we always take the RMSE from a reconstruction over two seconds. For the final implementation we plan an combination of P-REx with a classical fringe tracker, where the fringe tracker frequency is reduced to a few Hertz. For this implementation it would then be enough to stabilize the fringes over several hundred microseconds. We therefore use the timespan of two seconds as an upper limit for the time over which it should give good results. However, also a shorter timespan of stable fringes would be good enough in order to use P-REx in combination with a low frequency fringe tracking system. 

The results of the piston reconstruction slightly depend on the actual used atmosphere for the simulation, as well as on the wind direction. In order to quantify the quality of the two algorithms, we ran the algorithm for 100 times with the same AO system but with a different phase screen for each run and a random wind direction. The mean result from these 100 runs is a RMSE of \SI{0.08\pm0.04}{\micro\meter}. From this result and \autoref{fig:recpiston} one can conclude that the P-REx method is, in this simple case with a good AO system and a single layer atmosphere, able to reconstruct the piston drift evolution over a single telescope. This confirms the P-REx concept and the tip-tilt approximation.

\subsection{Parameter Study}
For the previous tests a comparably good AO system was used, in order to test the algorithm under ideal conditions. However, there are several aspects which could possibly decrease the quality of the piston reconstruction. For example the number of WFS sup-apertures and DM actuators was set to a comparably high value, in order to get a good spatial sampling. But real AO systems use typically a lower sampling. Other properties which could possibly play a role are the brightness of the guide star, the amount of turbulence in the atmosphere, usually represented by the value of D/r$_0$, the wind speed, or the used time average. We ran different simulations, in order to test the dependency of the results from these parameters. As all these simulations are still under the assumption of a ideal single layer atmosphere, the important result is not so much the actual performance as this will decrease for a realistic atmosphere. More important for us is first to understand which parameter of the atmosphere and the AO system have an important influence on the performance of P-REx.

One parameter which has no direct importance on the result of the piston reconstruction is the wavelength of the AO system. However, the science wavelength has a large impact on the interpretation of the results.  As the fringe contrast in an interferometer relies on the difference in phases between the two combined light beams, the phase error has to be smaller than a certain fraction of the wavelength. The acceptable piston reconstruction error therefore depends on the science wavelength. This means, that for smaller wavelengths the piston error has to be smaller than for larger ones. In \autoref{tab:piston}, three different error regimes for three different wavelength regimes are summarized. The error regimes are twice the wavelength below which the fringes become visible, a quarter of the wavelength where one starts to see stable fringes and a tenth of the wavelength where the fringe contrast stays stable. These regimes apply for the piston difference of an interferometric baseline. In order to use them on the piston drift at a single telescope, as we want to do it, a factor of $\sqrt{2}$ has to be applied.

\begin{table}
	\centering
	\caption{Uncertainty ranges for the piston reconstruction in different wavelengths.}
	\label{tab:piston}
	\begin{tabular}{rccc}
		\toprule
		& optical & NIR & MIR \\
		& 0.5 \si{\micro\meter} & 2.0 \si{\micro\meter} & 10.0 \si{\micro\meter} \\
		\cmidrule(lr{5pt}){1-4}
		2$\lambda$   & 0.71 \si{\micro\meter} &
		2.83 \si{\micro\meter} & 14.1 \si{\micro\meter}\\
		$\lambda$/4  & 0.08 \si{\micro\meter} &
		0.35 \si{\micro\meter} & 1.77 \si{\micro\meter} \\
		$\lambda$/10 & 0.04 \si{\micro\meter} &
		0.14 \si{\micro\meter} & 0.70 \si{\micro\meter}\\
		\bottomrule
	\end{tabular}
\end{table}

However, it is not very helpful to have nine different cases to analyze. In order to avoid this, the error budget can be summarized in four cases, which define the usability of P-REx:
\begin{itemize}
	\item $\Delta$P $<$ 0.05 \si{\micro\meter}: Piston reconstruction works for all wavelengths
	\item $\Delta$P $\approx$ 0.12 \si{\micro\meter}: works very good in the NIR and MIR and fairly well in the optical
	\item $\Delta$P $\approx$ 0.4 \si{\micro\meter}: works very good in the MIR, reasonable well in the NIR, and is not usable to co-phase in the optical
    \item $\Delta$P $\approx$ 1.5 \si{\micro\meter}: system only usable in MIR
\end{itemize}
With this scale one can now determine the performance of the P-REx algorithm under different conditions. This will again be applied on the RMSE over two seconds, as this is the time range where the algorithm should be able to work alone, without the additional fringe tracker.

We now started to run simulations with varying parameters of the AO system and the atmosphere, in order to understand the effects on the piston reconstruction. All the simulations are still under the assumption of a single layer atmosphere. A more realistic multilayer atmosphere is discussed in the next section. During the tests we found that most of the parameters are not critical as long as they stay in a reasonable range and the AO system works properly. This was for example shown for the atmospheric turbulence, which has only little effect as long as r$_0$ is in a range from 15 to 40 \si{\centi\meter}. The same goes for the wind speed. Another test was the luminosity of the guide star, which has no effect as long it is bright enough for the AO system to work. 

\begin{figure}
	\includegraphics[width=0.97\columnwidth]{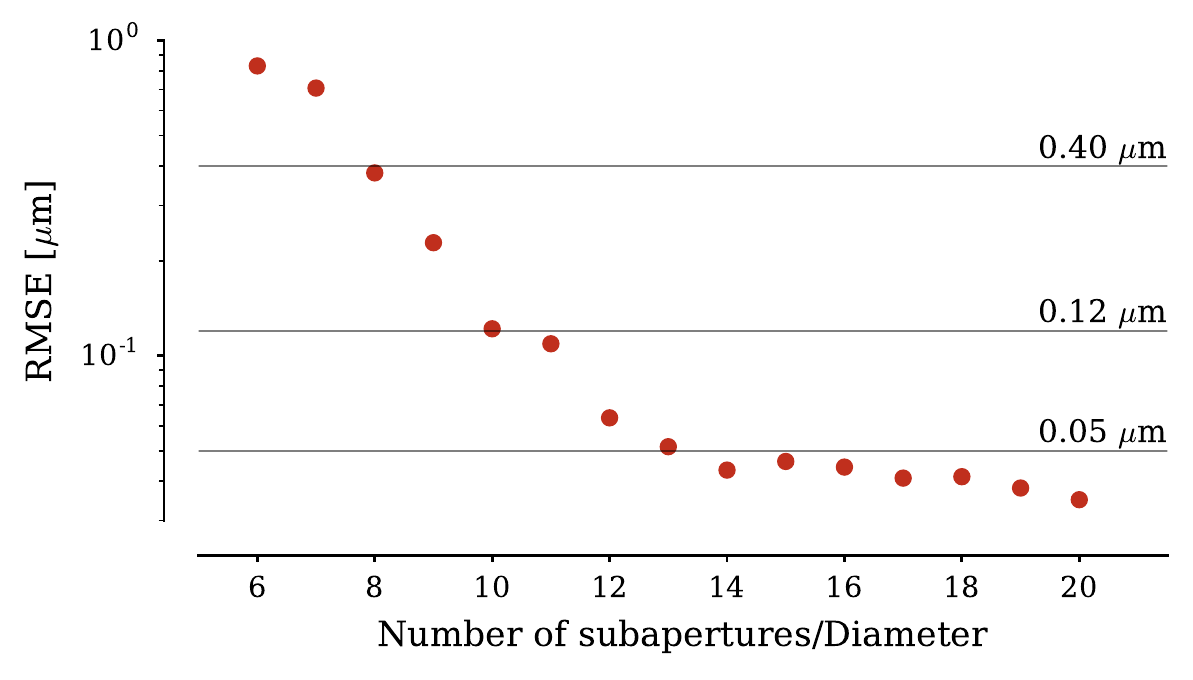}
	\caption{Dependency of the piston reconstruction on the spatial sampling. The plot shows the RMSE dependent on the number of WFS measurements in one telescope diameter. The gray lines indicate the three error regimes as discussed in the text.}
	\label{fig:dep_sampling}
\end{figure}

A factor which has an effect is the number of sub-apertures of the wavefront sensor. The spatial sampling of the WFS has two different effects on the results of the P-REx algorithm. The first one is simply that a higher number of sensors allows a more accurate measurement of the wind velocity. This is also based on the result that the error of the wind detection scales with the size of the WFS sub-apertures. The second effect is that the strength of the turbulence scales with r$_0$. Therefore, one could also test how good the WFS sampling per r$_0$ has to be for a working system. But as both effects tend into the same direction (more measurements mean a better result), it is  at this point not really important which effect dominates. This is however only the case as long as the sampling is not unreasonable high, which is usually given from the dimensioning of the AO system. \autoref{fig:dep_sampling} shows the expected behavior: With an increasing number of measurements per diameter the error of the piston reconstruction decreases. From the simulations we conclude that at least ten WFS sub-apertures per telescope diameter (meaning a 10x10 WFS) should be available in order not to restrict the piston reconstruction. This means that the AO systems at the VLTI UTs \citep[CIAO/MACAO,][]{Arsenault2003,Kendrew2012} and at the LBT \citep[LBT-FLAO,][]{Esposito2011} are plausible candidates for the piston reconstruction with P-REx. Especially the LBT-FLAO  should deliver good results, as the pyramid sensor has a WFS resolution of 30x30 measurements in the lowest binning mode. Our simulation showed a similar effect for variations of the telescope diameter. The performance of P-REx tends to decrease for smaller telescopes, as the area that is used to measure the ground layer turbulence gets smaller and with that also the accuracy of the cross-correlation decreases. However, this is more difficult to quantify as the AO systems are designed for each special case and therefore for example the WFS sampling is adapted to the telescope size and the expected D/r$_0$. As all these parameters have an influence on the result of the simulation this is quantified in more detail in the next subsection, where we take a more detailed look on existing AO systems.

\subsection{Performance for existing SCAO systems}
\begin{figure}
	\includegraphics[width=0.97\columnwidth]{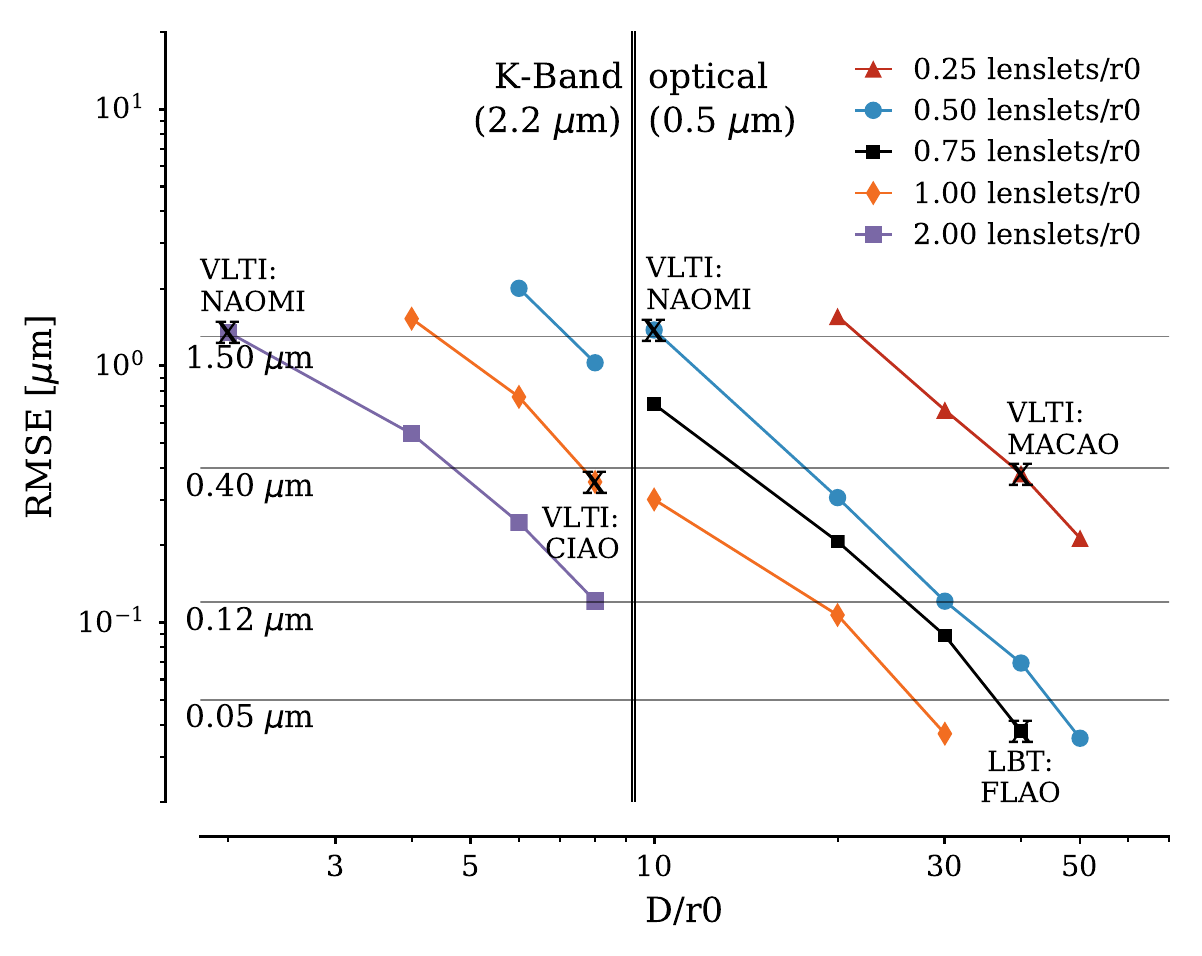}
	\caption{P-REx performance for a varying D/r$_0$ and different samplings of the WFS. The plot shows results in the K-Band (left part) and in the optical (right part). The different samplings per r$_0$ are shown in different colors, as indicated in the legend. The gray lines are the same error regimes as before.}
	\label{fig:superplot}
\end{figure}

With the previous findings the performance of the piston reconstruction for simulations of different set-ups of AO systems can be tested. This is done in order to test which of the existing AO systems is usable for P-REx. The results are shown in \autoref{fig:superplot}, where the error of the piston reconstruction is shown as a function of D/r$_0$.  For the plot a good seeing of \SI{0.5}{\arcsecond} was used, which corresponds to a r$_0$ of \SI{20}{\centi\meter} in the optical and \SI{1}{\meter} in the K-Band. The NAOMI system from VLTI \citep{Dorn2014} is shown in both wavelengths, as it works at \SI{1.65}{\micro\meter} in between the shown wavelengths. All the results in this plot are the average from three runs with different phase screens, in order to avoid an influence of any specific phase screen.

As we used a constant seeing for this test, the D/r$_0$ is varied by changing the diameter of the used telescope. This leads to the first conclusion that the performance of P-REx decreases with a decreasing telescope diameter, shown in the figure by lines of constant sampling. Therefore, larger telescopes are better suited for the application of the system. This result gets even worse for the \SI{2}{\meter} telescopes (ATs) of the VLTI, as the NAOMI WFSs use arrays of 4x4 lenses and therefore have a very low sampling. This further decreases the result as the lines from different sampling in \autoref{fig:superplot} and also \autoref{fig:dep_sampling} clearly show. Since in the near future NAOMI will be the only AO system for the ATs and the result is in general best for larger telescopes, we now focus on the \SI{8}{m} class telescopes.

For the UTs at the VLTI there are currently two AO systems with a similar WFS sampling. These are the CIAO system in the NIR with a 9x9 SH-WFS and the MACAO system in the optical with a 60 element curvature WFS. Both systems deliver P-REx results in the order of \SI{0.3}{\micro\meter}. These results are therefore in the regime where P-REx is usable in the NIR and works very well in the MIR. The last available system is FLAO at the LBT. With a maximum sampling of 30x30 for the pyramid sensor, this system has the best preconditions for the use of P-REx, which is also shown in the simulation.  With RMSE values of around \SI{0.06}{\micro\meter}, P-REx should work very well in the NIR and should also deliver good results in the optical. So far all the test were done with good seeing conditions and a single layer atmosphere. The so far mentioned performance is rather an upper limit, but the test already show for which instruments P-REx is best suited and what results could theoretically be achieved.

\subsection{Multilayer Atmosphere}\label{se:multi}
\begin{table}
	\caption{Composition of a simulated multilayer atmosphere as it is used in this work.}
	\label{tab:atm}
	\centering
	\begin{tabular}{llcccc}
		\toprule
		Layer & \# & Altitude & Fraction & Wind & Wind\\
		& & & & speed & direction \\
		& & m & \% & \si{\meter\per\second} & $^\circ$ \\
		\cmidrule(lr{5pt}){1-6}
		Ground 		 	& 1 &  0    & 45 & 10 &  0  \\
		& 2 &  400  & 13 & 12 &  5  \\
		& 3 &  1000 & 11 & 10 & -5  \\
		Medium 			& 4 &  1800 &  9 &  8 & -10 \\    
		& 5 &  2500 &  6 &  6 & -15 \\
		& 6 &  5000 &  5 & 10 & -5 \\
		High			& 7 &  8000 &  4 & 20 & 15 \\
		& 8 & 11000 &  4 & 25 & 25 \\
		& 9 & 15000 &  3 & 15 & 30 \\
		\bottomrule
	\end{tabular}
\end{table}

In our previous simulations only a single atmospheric layer has been used. For the next tests we used a realistic multilayer atmosphere, whose composition is shown in \autoref{tab:atm}. The strength of the different layers is based on different studies, mainly on \citet{Avila2004} and \citet{Andersen2006}. A similar composition has been found for the LBT \citep{Egner2006,Masciadri2010} and for the VLT \citep{Clenet2010}. An important point for this work is also the wind direction in the different layers. \citet{Avila2006} used data from San Pedro Martir in Mexico and found that in general the wind direction varies very little with the altitude. In the majority of their observing nights the wind direction stayed within approximately 60 degrees and showed no systematic behavior with increasing altitude. These findings agree with older results \citep{Schoeck1998,Gentry2000}.

\begin{table}
	\centering
	\caption{Wind measurements for different AO systems with single and multilayer atmosphere. The theoretical wind speed of the ground layer is \SI{10}{\meter\per\second}. The RMSE is, as usual, taken over a two second measurement.}
	\label{tab:atm_res}
	\begin{tabular}{lccc}
		\toprule
		AO system	& Single or 	& Wind 			& RMSE\\
					& multilayer	& measurement	&\\
					&	& \si{\meter\per\second} & \si{\micro\meter}\\
		\cmidrule(lr{5pt}){1-4}
		SCAO	& single	& \num{9,56 \pm 0,45}	& \num{0.066} \\
				& multi		& \num{9,42 \pm 0,85}	& \num{0.426} \\
		GLAO	& multi		& \num{9,41 \pm 0,63}	& \num{0.195} \\
		\bottomrule
	\end{tabular}
\end{table}

The performance of P-Rex with this atmosphere was then tested with a single-conjugated adaptive optics system (SCAO) and with a ground layer adaptive optics system (GLAO). As results of these tests, the measured wind  velocity and the RMSE are shown in \autoref{tab:atm_res}. As the ground layer is the dominant layer in the atmosphere, the detected shift in the cross-correlation is from the ground layer and the measured wind is the ground layer wind. The multilayer atmosphere effects the results of P-REx in two different ways: Firstly, it reduces the accuracy of the wind velocity measurement. The achieved values in \autoref{tab:atm_res} show a similar wind measurement for a single- and multilayer atmosphere, but the error of the multilayer wind measurement is twice as big as the error from a single layer. This is simply due to the other layers that act as an additional noise source in the cross-correlation. The second error source of the multilayer atmosphere is that the cross-correlation detects the dominant wind vector from the ground layer, but the tip and tilt are measured over the whole atmosphere. Therefore, P-REx calculates the product of the ground layer wind with the whole tip and tilt. This is not a huge problem as long as the layers do not move in opposite directions. As atmospheric studies have shown that the wind direction stays fairly stable with increasing height \citep{Gentry2000,Avila2006}, this is a not a restriction here. However, it does introduce an additional error in the final piston reconstruction. Our simulations have shown, that the piston reconstruction works properly as long as a minimum of \SI{65}{\percent} of the turbulence are located in the ground layer. The results in \autoref{tab:atm_res} show that the quality of the piston reconstruction clearly decreases with a multilayer simulation. The RMSE from the multilayer simulation is up to a factor of 6  bigger then in single layer simulations. Nevertheless, the value is still in an area where it is good enough for observations in the infrared. This also shows, that in order to get a better understanding of the effects from a multilayer atmosphere it is necessary to test the system on real data, as a realistic atmosphere cannot be perfectly modeled.

\begin{figure}
	\includegraphics[width=0.97\columnwidth]{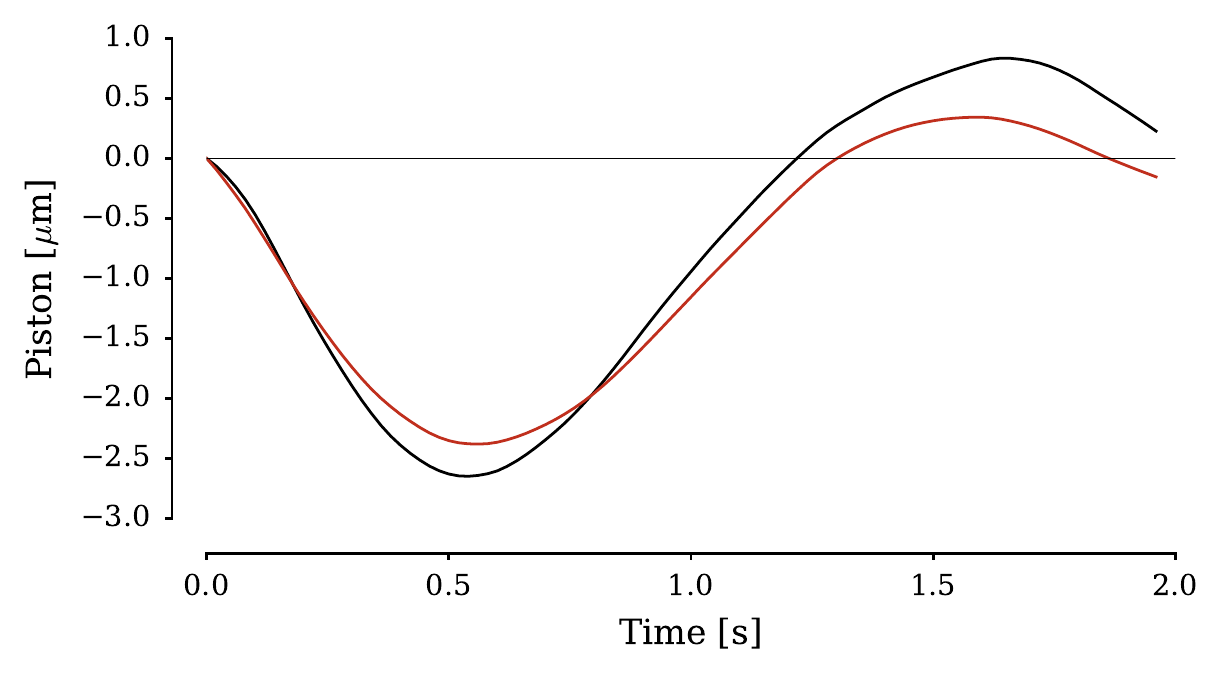}
	\caption{Piston reconstruction for a GLAO system  with a multilayer atmosphere. The theoretical piston is shown in black and the reconstructed piston is shown in red.}
	\label{fig:glao_full}
\end{figure}

We run a Final simulation to test the results from a multilayer atmosphere with a laser guide star (LGS) GLAO system.  The idea for the GLAO system is that one can get a better wind vector measurement from the data, as the noise due to the upper layers of the atmosphere is reduced. An important point for the piston reconstruction is that one has to use the wind vector from the ground layer, but the tip and tilt from the whole atmosphere, as one would otherwise only detect the ground layer piston evolution. This approach works, as the different atmospheric layers are expected to move in similar directions, as discussed earlier. This approach is also realistic, as the measurement from the LGS does not include the tip and tilt modes. Therefore the tip and tilt have to be measured with a natural guide star, which always gives values for the whole atmosphere. Consequently, the approach with the complete tip and tilt values gives better results and is also more practical for a real use of the system. The results in \autoref{tab:atm_res} show that the wind measurement and also the piston drift reconstruction significantly improve in comparison to a SCAO system. An example for the piston reconstruction with a GLAO system is shown in \autoref{fig:glao_full}

\subsection{Results from Simulations}
In conclusion, the simulations show promising results for the piston reconstruction. The key results are the following:
\begin{itemize}
	\item The piston reconstruction works well for a single layer atmosphere under typical seeing conditions.
	\item The most important criterion is to have a sufficiently high WFS sampling, typically larger than 10x10 sub-apertures with an individual sub-aperture size in the order of r$_0$. The higher the sampling, the more precise  is the wind measurement. 
	\item The best results for realistic systems were achieved for an \SI{8}{\meter} telescope with a 30x30 element WFS. This is given for example by the LBT-FLAO system, but also the CIAO and MACAO systems at VLTI showed promising results.
	\item In order to get a reliable wind vector, the majority of the turbulence has to be located in the ground layer, which is usually the case. The ground layer wind vector can be determined also from a realistic multilayer atmosphere with high layer turbulence. The simulations showed a usability of P-REx especially at infrared wavelengths, even with a complex multilayer atmosphere.
	\item The use of a GLAO system can further improve the result in comparison to a SCAO system, due to the cleaner wind measurement of the dominating ground layer.   
\end{itemize}
The only effect that was not considered yet in the simulation, is the temporal evolution of the atmosphere beyond the assumption of TFFH, namely the boiling. There are possible ways to include boiling into the simulations \citep[see e.g.][]{Assemat2006,Berdja2007}. However, the main parameters of boiling, such as the strength, are not clear and simulating a phase screen with chaotic behavior is not well understood. Therefore, we now use on-sky data to better understand further effects.

\section{LBT Data}\label{sec:flao}
\begin{table}
	\centering
	\caption{Main properties of the used data. The seeing and the wind velocity are averaged over the datasets, measured at the LBT.}
	\label{tab:flaodata}
	\begin{tabular}{lcc}
		\toprule
		 & Night 1 & Night 2 \\
		\cmidrule(lr{5pt}){1-3}
		Date & 17.9.2012 &  20.9.2012 \\
		Number of datasets & 12 & 5 \\
		Average wind velocity & \SI{2.31}{\meter\per\second} & \SI{4.51}{\meter\per\second}\\
		Average seeing & \SI{0.50}{\arcsecond}	& \SI{0.90}{\arcsecond}  \\
		Wavelength & \SI{650}{\nano\meter}	& \SI{650}{\nano\meter} \\
		r$_0$& \SI{27}{\centi\meter} & \SI{37}{\centi\meter} \\
		D/r$_0$(D=\SI{8}{\meter})	& 30 & 22 \\
		D/r$_0$(D=\SI{3.5}{\meter})	& 13 & 9 \\
		\bottomrule
	\end{tabular}
\end{table}

In 	order to better understand the effects of real atmosphere dynamics on the piston reconstruction, we now report on testing the algorithms on on-sky data from the LBT-FLAO system. FLAO uses a pyramid sensor which has three different binning modes \citep[see e.g.][]{Esposito2010,Esposito2011}. The used data were taken in the lowest binning mode, which means that the WFS measurements consist of 30x30 slopes. The data consist of 17 datasets with measurements over four seconds. They were taken at two different nights in 2012, with a very good seeing in the first night and medium conditions in the second one. The main parameters of the datasets are listed in \autoref{tab:flaodata}.  All data are from the LBT-SX telescope and were kindly provided to us by A. Puglisi and S. Esposito in personal communication. 

\subsection{Wind Measurement}
\begin{figure*}
	\includegraphics[width=0.9\textwidth]{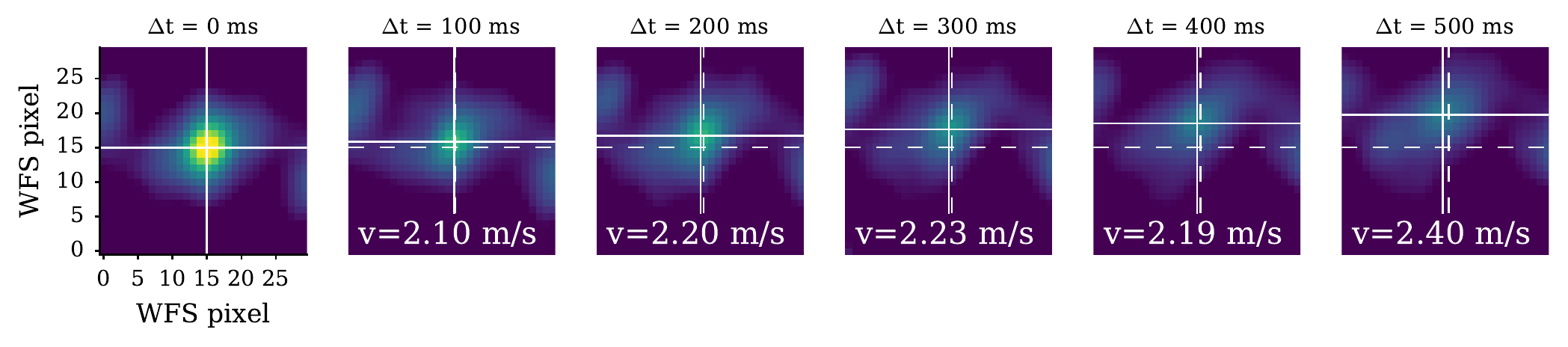}
	\caption{Wind measurement from FLAO data. The elapsed time in the cross-correlation is increasing from the top left to the bottom right with steps of \SI{100}{\milli\second} per image. The center of the image is shown with a dashed cross and the peak of the image with a solid cross. The measured wind velocity is given for each image.}
	\label{fig:wind_flao}
\end{figure*}
\begin{figure}
	\includegraphics[width=0.97\columnwidth]{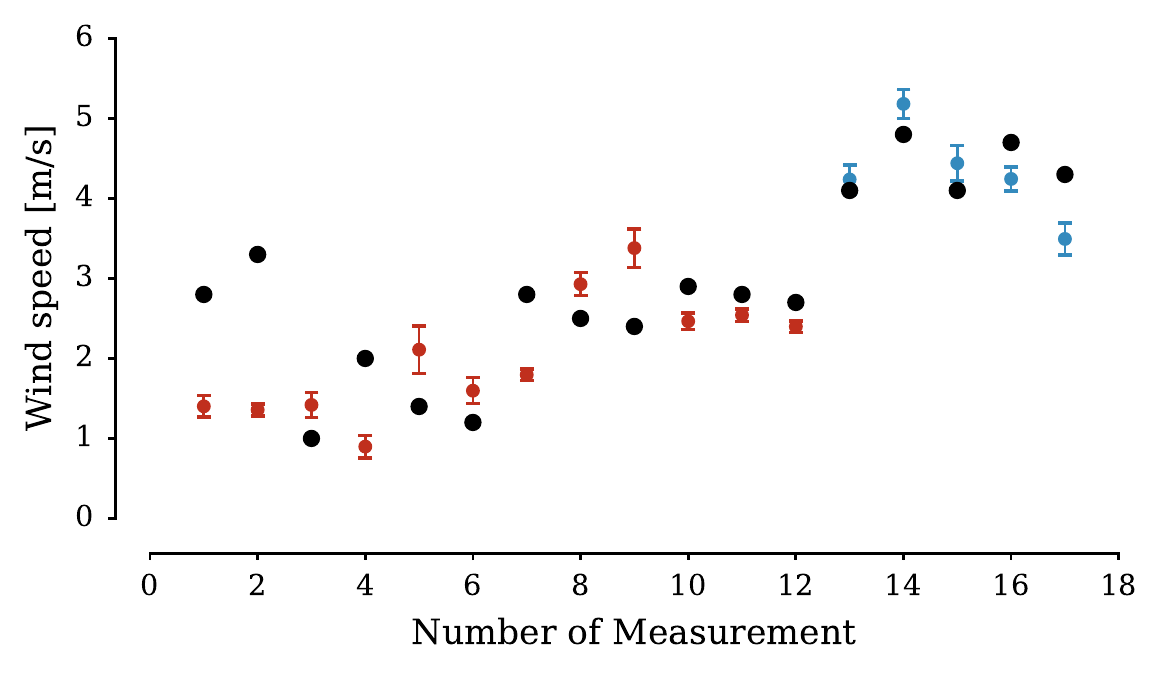}
	\caption{Comparison of the wind measurement from the cross correlation (in red from the first night, in blue from the second one) with the values from the LBT wind sensor. The error bars at the cross correlation data are the standard deviation over each dataset.}
	\label{fig:wind_flao_comp}
\end{figure}

A first test of the LBT data has the goal to see whether it is possible to measure a dominant wind vector consistently. This is a crucial measurement, as the wind vector is absolutely necessary for the use of P-REx. \autoref{fig:wind_flao} shows the cross-correlation of the POL slope data for different time shifts, taken from one single dataset. It is clearly visible that the peak position shows an increasing distance from the center of the image, as it is expected for an increasing time difference. From the peak position the wind velocity can directly be calculated, under the assumption that the ground layer is the dominant layer. The wind velocity stays nearly constant over the measurement of half a second (\autoref{fig:wind_flao}). The results can be verified by comparing them to the wind measured directly at the LBT (\autoref{fig:wind_flao_comp}). The main trends in the data, such as a higher wind velocity in the second night, are visible in both measurements. There is a certain difference between the two measurements, which can be explained with the fact that the sensor at the LBT only measures the wind velocity directly at the telescope, while the cross-correlation sees the whole atmosphere. In total the wind measurement from the cross-correlations seams reliable. 

\subsubsection{Boiling}
The main difference between the real data and the simulated one is that the intensity of the correlation peak in the simulation stays constant, while the peak intensity in  \autoref{fig:wind_flao} clearly decreases. This is expected, as the real atmosphere does not only show a translation but also changes due to boiling, which reduces the intensity of the correlation peak. The decreasing peak value can therefore be used to verify the intensity of boiling.

\begin{figure}
	\centering
	\includegraphics[width=0.6\columnwidth]{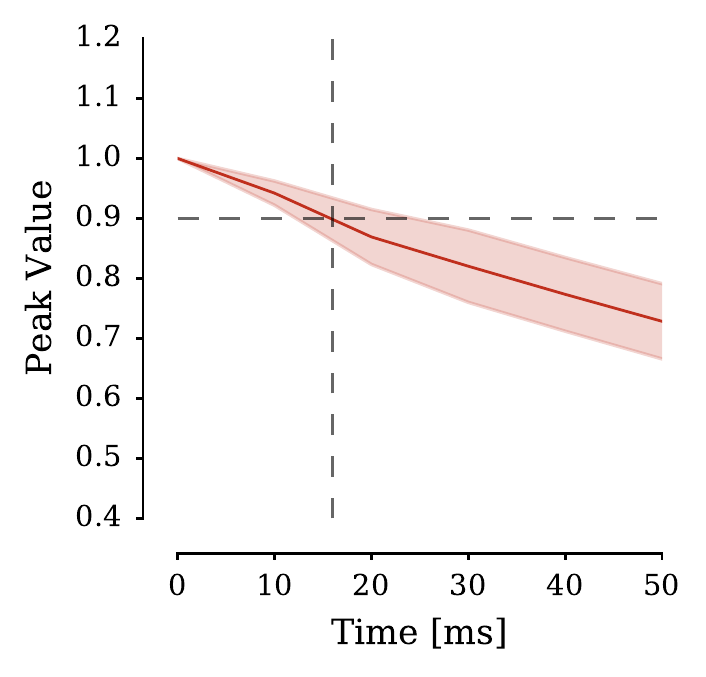}
	\caption{Evolution of the mean correlation peak over time as a red line. The shaded area shows the standard deviation of the data. The dashed lines show the time at which the peak is at \SI{90}{\percent} of its initial value.}
	\label{fig:flaopeak}
\end{figure}

For an atmosphere solely moving by the wind velocity in perfect frozen flow, the intensity of the cross-correlation peak would stay constant at one. By mapping the intensity of the peak over time it can be measured to what degree TFFH is valid in the actual data. The decrease in peak intensity in real data is due to boiling, but as well due to the different layers of atmosphere moving into different directions. The evolution of the cross-correlation peak intensity is shown in \autoref{fig:flaopeak} as the mean value from all of the FLAO datasets. The peak intensity decreases over time, as it is expected. A value to quantify the impact of TFFH is $t_{90}$, the time over which TFFH is responsible for \SI{90}{\percent} of the atmospheric evolution. From these data, we can conclude that $t_{90}$ is approximately \SI{16}{\milli\second}. A similar study was done by  \citet{Schoeck2000}, with data from the Starfire Optical Range in Albuquerque, New Mexico. They found slightly higher values with an average $t_{90}$ of \SI{25}{\milli\second}. These results furthermore confirm the assumption that TFFH is correctly describing the temporal atmospheric phase evolution over short timescales and verify the used timescales of \SIrange[range-units=single]{10}{20}{\milli\second} for the wind vector estimation. Our  analysis shows that boiling is not a limiting factor for the usage of P-REx: The wind vector measurement can be done on timescales where boiling is clearly present, without loosing sensitivity. This is shown in \autoref{fig:wind_flao}, where the correlation peak is visible over timescales of \SI{0.5}{\second}. However, P-REx has to be applied on much shorter timescales. The reasons for this are that the piston variations occur on much shorter timescales and that the importance of the frozen flow hypothesis, which is the basis ob the piston calculation, decreases at larger timescales. The best solution is therefore to decouple the wind measurement and the piston drift reconstruction. With this, it is easily possible to run P-REx efficiently on timescales in the order of \SI{10}{\milli\second}, where the frozen flow dominates the atmospheric evolution. On the same time the wind measurement is taken over longer timescales, which decreases the error in the wind vector.

\subsection{Single Aperture Method}
\begin{figure}
	\includegraphics[width=0.95\columnwidth]{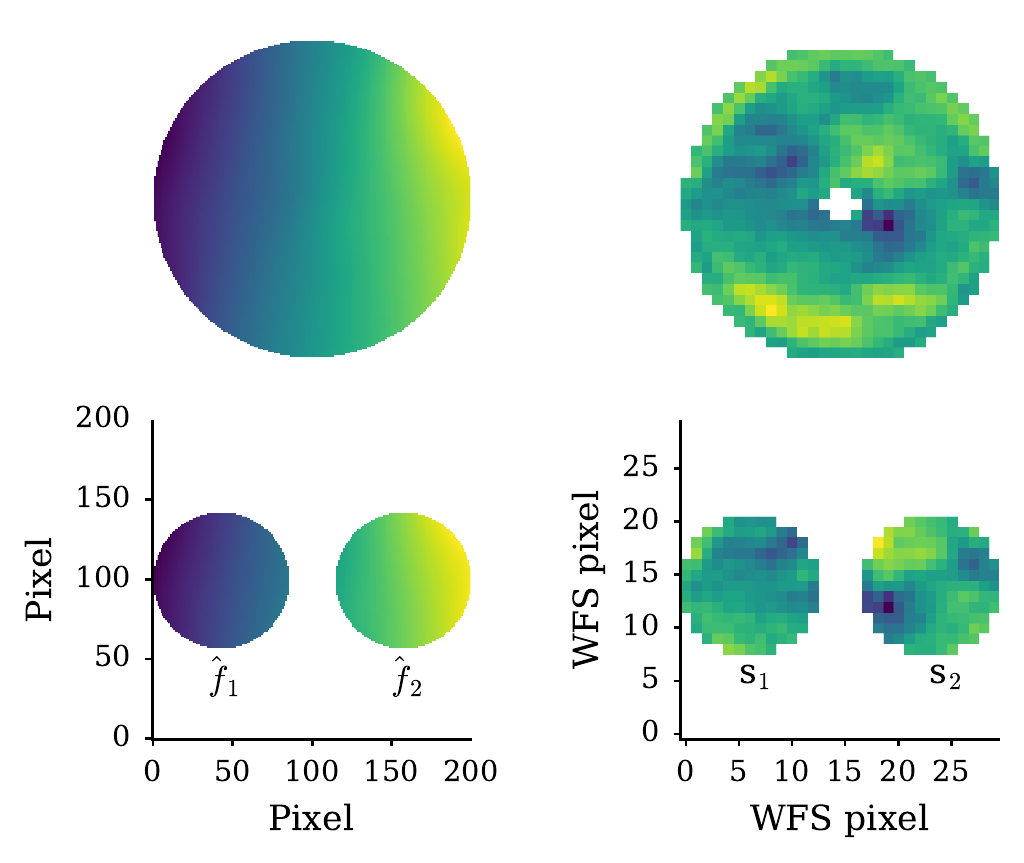}
	\caption{Explanation of the single aperture mode with the two subapertures from the reconstruction (left) and from the slope data (right).}
	\label{fig:sa}
\end{figure}

We now want to test the P-REx algorithms on the LBT data. In contrast to the simulations we now do not have the theoretical values for the piston evolution. Ideally, one has to test the P-REx algorithm on interferometric data and compare it to the fringe tracker measurement from the same dataset. However, as we do not have access to interferometric data we will test the usability of P-REx on single aperture data. In order to do so, we applied the following  verification method. We define two circular subapertures containing each about a quarter of the actual slope data. The idea is then to think of these two regions as virtual telescopes and to calculate the difference in piston evolution between them. This measurement can be compared to the difference in piston evolution for the same regions from a wavefront reconstructed over the full wavefront. The principle of this \textit{single aperture method} is illustrated in \autoref{fig:sa}. This means that from now on we are not working with a single telescope anymore, but are cutting out a virtual two-element interferometer from the FLAO data.

\subsubsection{Reconstruction}
Starting with the verification part of the single aperture method, the first step is to reconstruct the wavefront from the POL slopes. For the reconstruction of the wavefront a reconstructor is used in order to calculate the Zernike modes from the slopes. The wavefront is then calculated from these modes. As there is no piston information in the slopes, the reconstructed wavefront $\widehat{WF}$ differs from the real wavefront $WF$, as it is reduced by the piston $\langle WF \rangle$:
\begin{equation}
	\widehat{WF} = WF - \langle WF \rangle~~~ \Rightarrow ~~ P_{\widehat{WF}} = \langle \widehat{WF}  \rangle =0
\end{equation} 
In the following formulas, the notation with a hat, such as $\widehat{WF}$, always corresponds to the piston-subtracted value. As shown in the left part of \autoref{fig:sa}, the next step is to cut out two circular frames, $\hat{f}_1$ and $\hat{f}_2$, from the full aperture wavefront. The whole wavefront is piston-subtracted. However, as the overall piston is different to the piston in the two sub-frames, the two sub-apertures have an individual piston value:
\begin{equation}
	\hat{f}_i = f_i - \langle WF \rangle ~~~ \Rightarrow ~~~ P_i = \langle \hat{f}_i \rangle \neq 0
\end{equation}
The value which is now of interest is the piston difference between the two frames which is equivalent to the real piston difference between these two regions of the atmosphere, as it can be seen in the following calculation:
\begin{align}
	\begin{split}
		\mathit{dP}_\text{recon} & = P_1-P_2 = \langle \hat{f}_1 \rangle - \langle \hat{f}_2 \rangle \\
		& =\langle f_1 - \langle WF \rangle\rangle - \langle f_2 - \langle WF \rangle\rangle  = \langle f_1 \rangle - \langle f_2 \rangle
	\end{split}
\end{align}
This value is then called the reconstructed piston difference ($\mathit{dP}_\text{recon}$) and is, calculated for each time step, the reference value for this P-REx test.

\subsubsection{Slopes}
The actual measurement is then done on the POL slopes, as shown in the right part of \autoref{fig:sa}. Again, we cut out the same regions from the slope data, which are now named $s_1$ and $s_2$. For these two regions one can then calculate the differential piston with the P-REx algorithm (see \autoref{equ:prex}):
\begin{equation}
	\Delta P_i(t) = \left[\langle s_{ix}\rangle\cdot v_x + \langle s_{iy}\rangle\cdot v_y\right]\cdot\Delta t
\end{equation}
By adding up the differential piston for each of the two parts, one gets the piston evolution $P_{i,sl}$, which is correct except for a constant factor, the initial piston value $P_{i,t=0}$:
\begin{equation}
	P_{i,sl}(T) = \sum_{t=0}^T \Delta P_i(t) = P_i(T) - P_{i,t=0}
\end{equation}
The difference of these two piston evolutions is then the result of this part, the piston difference from the slope measurement $\mathit{dP}_\text{sl}$:
\begin{align}
	\begin{split}
		\mathit{dP}_\text{sl}(T) & = \sum_{t=0}^T \Delta P_1(t) - \sum_{t=0}^T \Delta P_2(t) \\
				& = P_{1,sl} - P_{2,sl} - \underbrace{(P_{1,t=0} - P_{2,t=0})}_{= C} = dP - C
	\end{split}
\end{align}
which is again the real piston difference between these two regions, except for the difference of the initial values.

\subsubsection{Combination}
The last step is to show that the measured values can be compared. From the fact that the reconstructed piston difference is the actual piston difference, one gets:
\begin{equation}
	\mathit{dP}_\text{recon}(t) = dP(t) = \mathit{dP}_\text{sl}(t) +C
\end{equation}
From the condition that $\mathit{dP}_\text{sl}(t=0) = 0$, this leads to: $ C = \mathit{dP}_\text{recon}(t=0)$ and with that to:
\begin{equation}
	\mathit{dP}_\text{recon}(t) - \mathit{dP}_\text{recon}(t=0) = \mathit{dP}_\text{sl}(t)
\end{equation}
This means that by subtracting the initial value from $\mathit{dP}_\text{recon}$, the results of the two measurements are equal.

\subsection{Test on simulation}
\begin{figure}
	\includegraphics[width=0.97\columnwidth]{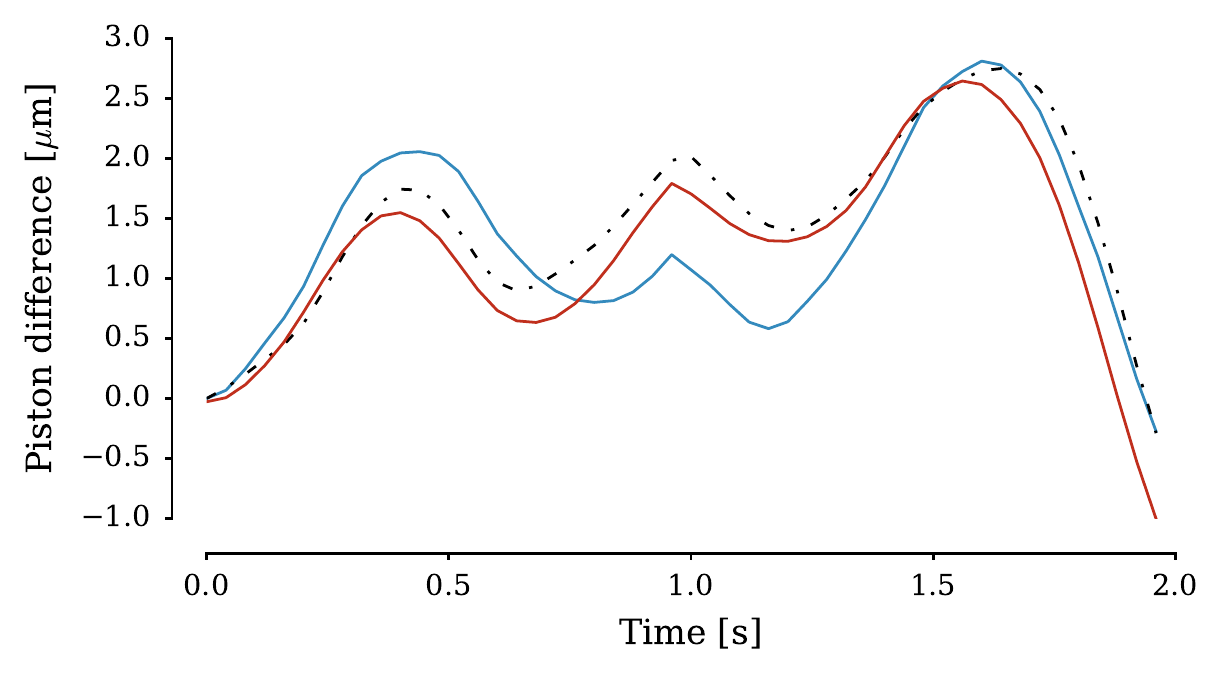}
	\caption{Single aperture method used on data from a simulation. The red line shows the piston difference calculated from P-REx, the blue line from the reconstructed wavefronts and the black dashed line the real values.}
	\label{fig:pistoncontrol_yao}
\end{figure}
The advantage of first testing the single aperture method on simulation data is that one can better understand the reliability and possible error sources of the method. We therefore run a simulation with an AO system comparable to FLAO. The single aperture method is applied to the simulation data exactly in the same way as it has to be done with the LBT data. The simulated data offers an additional verification step, as the input atmosphere is known. This is shown in \autoref{fig:pistoncontrol_yao}, where the piston difference from the slopes, from the reconstructed wavefront, and from the real atmosphere are shown.

The first thing to mention in \autoref{fig:pistoncontrol_yao} is that the values from the reconstructed wavefront and the real values are not equal. The difference between these two measurements is due to the fact that the reconstructed wavefronts are not error free. The wavefront reconstruction with Zernike modes lacks of small scale structure and includes a fitting error due to the used AO spatial resolution. This disagreement between the real atmosphere and the reconstructed wavefront is relevant here, as the missing small scale structure gets more important by cutting out sub regions from the big phase screen. This then leads to the mismatch between the two theoretical values in \autoref{fig:pistoncontrol_yao}. 

The error in the control loop leads to the result that the P-REx values from the single aperture method seem to be worse than they actually are. The RMSE between the P-REx values and the values from the reconstructed wavefront is \SI{0.47}{\micro\meter} for this simulation, while the RMSE between the P-REx values and the measurement from the real atmosphere is only \SI{0.34}{\micro\meter}. These results show that the single aperture method only gives a lower limit for the quality of the reconstructed piston, as the reconstructed reference  values are not perfectly correct. We can therefore expect that the true performance is systematically better than suggested by this test.

Furthermore, the RMSE of \SI{0.34}{\micro\meter} appears to be  worse than expected from previous simulations. This is mainly due to the fact that the P-REx results improve with the sampling of the WFS and the size of the telescope, as shown in \autoref{fig:superplot}. By cutting out the two regions, the sampling decreases from one 30x30 WFS to two 12x12 WFS with a telescope size of \SI{3.8}{\meter} for each telescope. For the used r$_0$ of \SI{20}{\centi\meter} this leads to a D/r$_0$ of 17.5 with a sampling of 0.75 lenslets/r$_0$. When one looks at these values in \autoref{fig:superplot} the result should be a RMSE in the order of \SI{0.4}{\micro\meter}. This shows that the results from this sections fit very well to previous simulations and also that the result should be better when one uses the whole telescope aperture with \SI{8}{\meter} diameter. Despite this limitation of our verification method, it is still usable on the LBT data to give an indication whether P-REx can work under real conditions.

\subsection{Results from LBT data}
\begin{figure*}
	\includegraphics[width=0.9\textwidth]{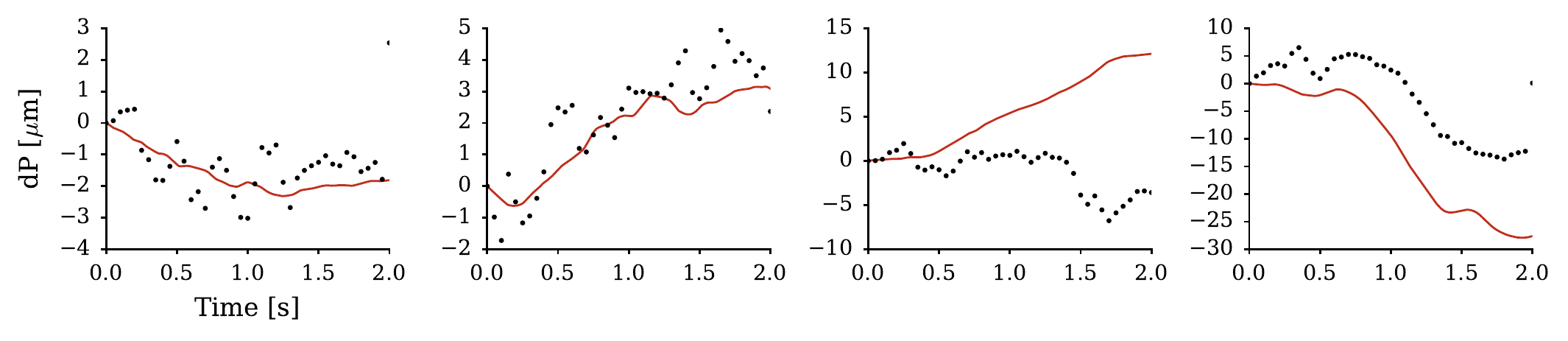}
	\caption{Four examples from the results of the single aperture method for LBT FLAO data. In all plots the red line is from the P-REx measurement and the black dots are from the reconstructed wavefront. }
	\label{fig:pistoncontrol_flao}
\end{figure*}
The single aperture method is now applied to the FLAO data. As mentioned earlier, the wind measurement needs a relatively long average in order to give consistent results. Therefore, the wind is measured over one second with a moving average. The P-REx calculation is then not limited by the wind measurement and can basically be done with the same frequency as the wavefront sensing. In order to improve the SNR of the measurement, an average over five tip and tilt measurements is taken, which corresponds to a timespan of \SI{5}{\milli\second}. This means that the P-REx algorithm would run with approximately \SI{200}{\hertz} in this specific case.

The results from the data are non-uniform. Some datasets show a very good agreement in the single aperture method, while in other datasets the two measurements show a very different behavior in the piston difference. This is shown for four examples in \autoref{fig:pistoncontrol_flao}. With this four examples one can see the full range of possible results. For the left two plots the reconstruction is very similar to the theoretical values over the whole time. The two right plots show two examples for a poor performance. While both of them lead to a high RMSE (5.9 and 8.0 \si{\micro\meter}), the actual trend is very different. While the reconstruction in the left plot is completely wrong, the reconstruction in the right plot shows the trend of the theoretical piston evolution rather good. However, a piling up error from the beginning decreases the final result. This is therefore an example where it becomes clear that a low frequency fringe tracker would very much improve the result on the P-REx feed-forward control residuals. 

\begin{figure}
	\includegraphics[width=0.97\columnwidth]{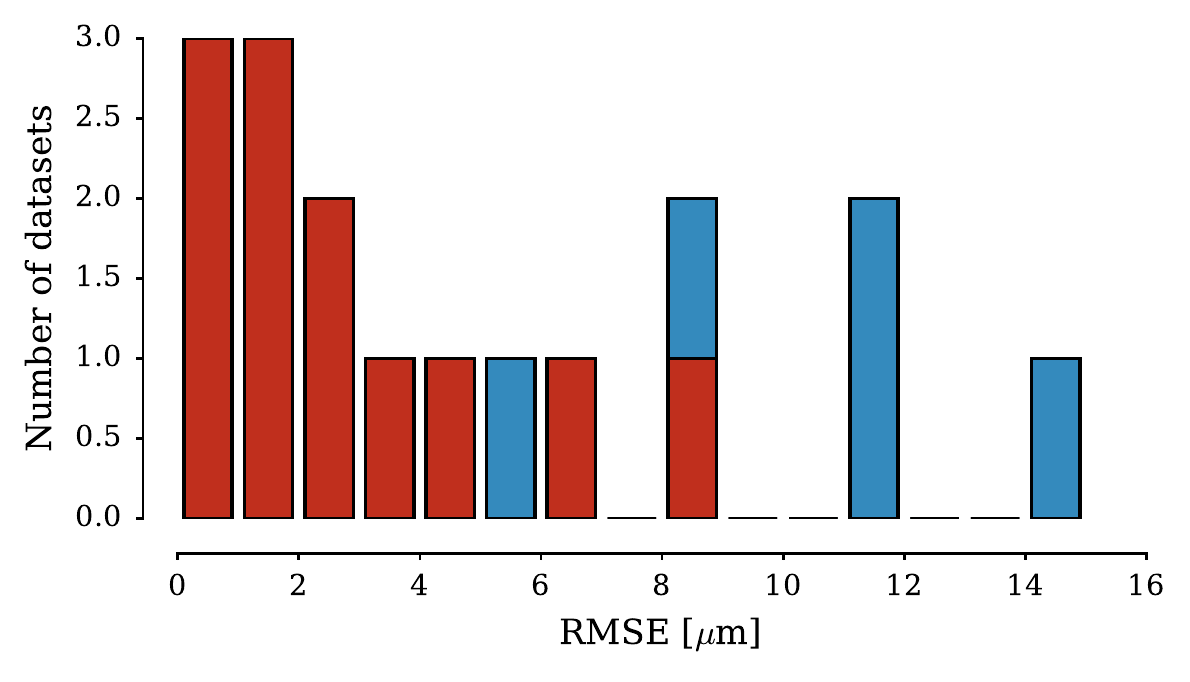}
	\caption{Histogram of the RMSE of the 12 FLAO datasets from the single aperture method. The datasets from the first night are shown in red and the datasets from the second night are shown in blue.}
	\label{fig:pistoncontrol_histo}
\end{figure}
The large range of different results shows that a conclusive statement from these data is challenging. In order to get some statistics from the data, \autoref{fig:pistoncontrol_histo} shows a histogram of the RMSE between the piston difference from P-REx results and from the reconstructed wavefront. A first conclusion from this histogram is that the results from the first night (shown in red, r$_0$ $\approx$ \SI{27}{cm}) are in general better than from the second night (shown in blue, r$_0$ $\approx$ \SI{37}{cm}). This shows that the piston reconstruction improves for better atmospheric conditions. In order to further understand the results, one needs to compare them to the expected values from simulations. \autoref{tab:flaodata} summarizes the atmospheric conditions during the observations. With these data we now use \autoref{fig:superplot} to analyze the expected quality of the piston reconstruction. As the single aperture method does only uses two sub-areas of the whole telescope, the size of the used area is roughly \SI{3.5}{\meter}. This leads to a D/r$_0$ of 13 for the first night and of 9 for the second night. For the present r$_0$, the wavefront sampling is in the order of 0.75 measurements per r$_0$. Looking at \autoref{fig:superplot}, a sampling of 0.75 measurements per r$_0$ for a D/r$_0$ of roughly 10 gives expected RMSE values in the order of \SI{1}{\micro\meter}. From the FLAO data half of the datasets from the first night have a RMSE below \SI{2}{\micro\meter} and two-thirds below \SI{3}{\micro\meter} (see \autoref{fig:pistoncontrol_histo}). In \autoref{tab:piston} we showed that in the MIR a piston error in the order of \SI{1.5}{\micro\meter} would be acceptable. When one considers that the results from the single aperture method are slightly to big due to the error in the wavefront reconstruction, the datasets with a RMSE below \SI{3}{\micro\meter} indicate a direct usability in the MIR.

In order to further interpret these results we have to consider different effects. On first view the RMSE values are much larger then expected from simulations, as the previous section showed that the single aperture method gives values for a single layer atmosphere in the order of \SI{0.5}{\micro\meter}. However, we found in \autoref{se:multi} that a turbulent multilayer atmosphere can decrease the results by up to a factor of six. The complex atmosphere in the on-sky data is therefore most likely the reason for the increased RMSE values, compared to the simulations of the single aperture method. Another point is that the reconstructed values are, as shown before, not perfectly correct. With these two effects, the complex atmosphere and the error from the reconstruction, taken into account, one can understand the discrepancy to the expected values from the simulation (RMSE $\approx$ \SI{0.4}{\micro\meter} for a \SI{3.8}{\meter} aperture and a single layer atmosphere, see \autoref{fig:superplot}). As expected the real atmosphere decreases the results of P-REx, but not worse than shown in simulations (\autoref{se:multi}) and not beyond the usability, at least in infrared wavelengths.

The comparison of the results from real data and from simulations leads to some conclusions. First of all, the results coincide
reasonably well with our simulations, when one takes the complex atmosphere into account. This confirms that the principal assumptions underlying the simulations are valid. This further approves that the boiling effect can be neglected over the here used timescales. Concluding that the seen effects in the simulations are valid, this then also means that the results from P-REx should significantly improve by using the whole telescope with a much higher D/r$_0$, as \autoref{fig:superplot} showed a major improvement towards the 8 m class telescopes. However, \autoref{fig:pistoncontrol_histo} also indicates that the results from the piston reconstruction depend on the atmospheric conditions and that P-REx is most usable under good seeing conditions, which is not unexpected.

\subsubsection{Atmospheric Conditions Adverse to P-REx}
\begin{figure}
	\begin{minipage}{0.97\columnwidth}
		\includegraphics[width=\textwidth]{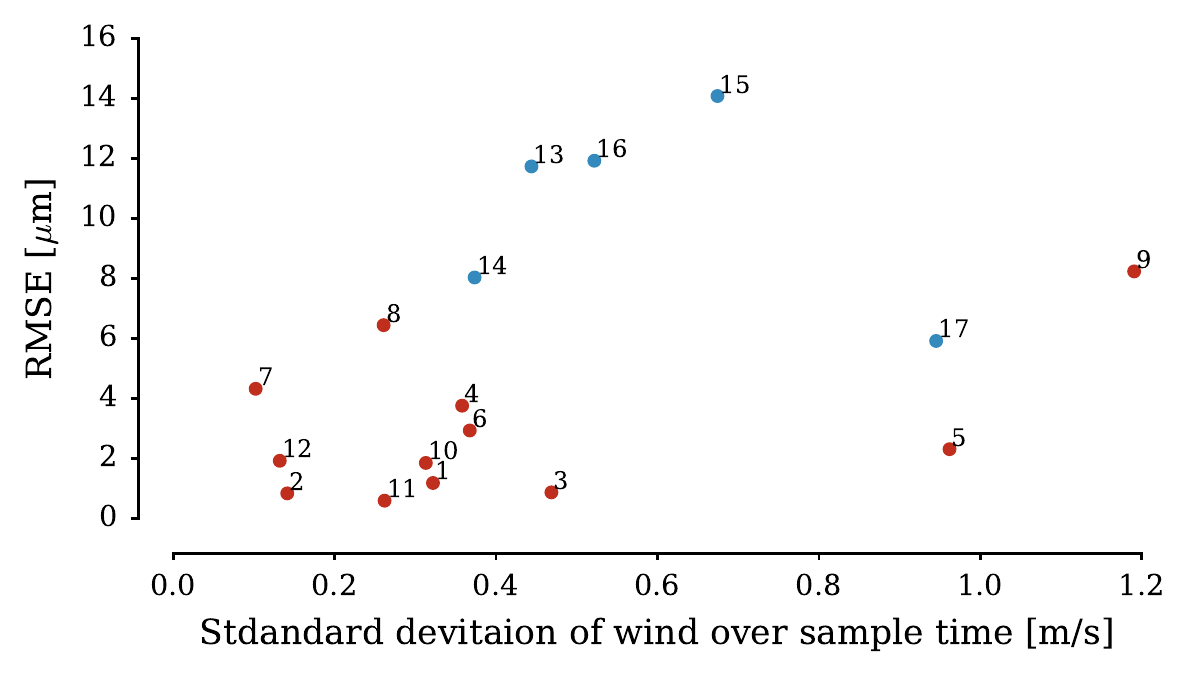}
	\end{minipage}
	\begin{minipage}{0.97\columnwidth}
		\includegraphics[width=\textwidth]{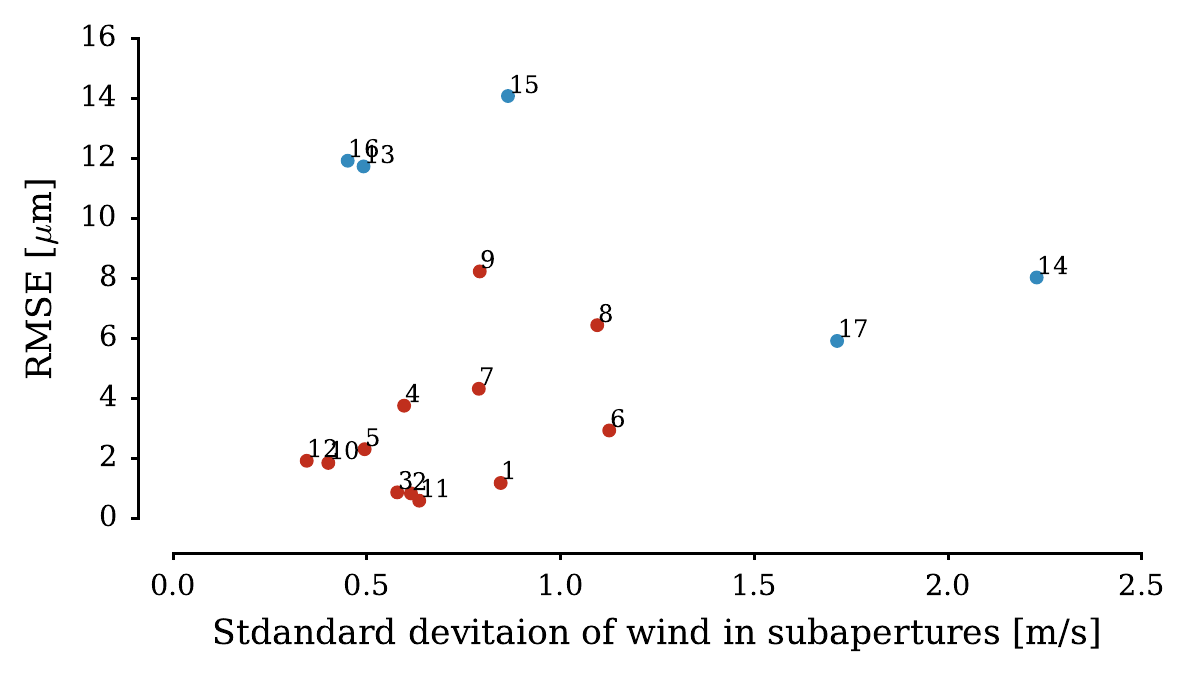}
	\end{minipage}
	\caption{Statistics of the single aperture mode with LBT FLAO data: the y axis in both images gives the RMSE value of the reconstruction with the single aperture method. The x axis of the upper image shows the standard deviation of the wind over the four seconds of the measurement. The lower image shows the standard deviation of the wind measurement in four sub apertures of the telescope. The number of the measurement is given for each data point in both plots with the number next to the data point. Data from the first night is shown in red, data from the second night is shown in blue.}
	\label{fig:pistoncontrol_rms}
\end{figure}

Despite some positive conclusions from the different datasets, the question remains why some of the data from the first night show comparably bad results from the piston reconstruction. The seeing conditions were not varying much during that night, but there are still some datasets which do not deliver good results. In order to look at different possibilities for that, \autoref{fig:pistoncontrol_rms} shows the RMSE from the different measurements as a function of the wind stability. This is done in two different ways: The first one is to study the temporal variance of the wind velocity, as shown in the upper plot of \autoref{fig:pistoncontrol_rms}, by plotting the RMSE against the standard deviation of the wind measurement over the four second dataset. The lower plot shows RMSE in comparison to the spatial variance of the wind velocity. This is done by measuring the wind in four different sub apertures and taking the standard deviation of these four measurements. This means that data points on the right side of the plots show an instable wind over time (upper plot) or over the area of the telescope aperture (lower plot). In order to better track the individual datasets, they are numbered from 1 to 17 and the number is printed next to the data point.

A first outcome of these plots is that the wind is much less stable in the second night. This was somehow expected, as the atmospheric conditions are in general worse in that night. Focusing on the data from the first night, one can see that the datasets with the biggest RMSE, which are datasets 8 and 9, show either a high temporal instability (9) or a high spatial variability (8). This shows that a bad result of the piston reconstruction can be the effect of a quickly changing or instable wind vector. However, this is not necessarily the case, as some data sets, such as number 5 and 6, show a high instability, but a good piston reconstruction. Therefore, the changing wind vector can be a performance limiting factor but does not necessarily has to be. For an actual implementation of P-REx one could think of a mechanism where P-REx is only used under good observing conditions and otherwise the usual fringe tracking approach is used. As shown here, the wind vector can be a possible test whether P-REx is usable. However, we do not have enough data right now to consider the details of such a system. We therefore refer to further test with a full interferometric dataset to investigate this.

\subsection{Conclusion from LBT data}
The tests with the single aperture method on LBT data largely confirm the results from the simulations. The main results from this part are the following:
\begin{itemize}
	\item A wind vector is clearly detectable in all the datasets, which is a prerequisite for the success of P-REx.
	\item The boiling intensity and timescale in these data are similar to previous studies. The found timescales confirm that boiling has only little impact on the P-REx performance.
	\item When taking all effects, such as the multilayer atmosphere and the decreased sensitivity due to a small aperture, into account, the performance of the piston drift reconstruction is comparable to the expectations from simulations. This suggests that the main assumptions in the simulations were valid and approves the results from simulations. One can therefore assume that the multilayer atmosphere has a decreasing effects on the results, but also that the P-REx performance for a full aperture of an \SI{8}{\meter} class interferometer is significantly better than for the here presented single aperture tests with the two small subregions.
	\item The P-REx tests on the FLAO data with good atmospheric conditions give mixed results for one third of the datasets. For the other two thirds, P-REx delivers an OPD RMSE below \SI{3}{\micro\meter}. When one considers the error due to the wavefront correction, this is even with the \SI{3.5}{\meter} aperture in the range where we assume a usability in the MIR.
	\item The tests showed that the results decrease for larger seeing (second night of the data). Furthermore, a varying wind vector and instable atmospheric conditions negatively influence the quality of the P-REx results, as it can be expected. This shows that one need to further analyze the conditions under which P-REx is usable.
	\end{itemize}
The most important points to take away from the tests on a single aperture are, that the atmosphere is, under good observing conditions, sufficiently well structured, stable and ground-layer dominated, in order to precisely measure an effective wind vector and apply the piston reconstruction algorithm. This leads to a principal beneficial usability of the P-REx algorithms for interferometric observations of faint targets.

\section{Discussion and Outlook}
With this work, we verified the assumptions leading to the P-REx concept design and its benefit for fringe tracking in realistic end-to-end simulations and real closed-loop AO data from the LBT. Extensive tests with data from simulated AO systems proved the capability of the piston reconstruction under typical atmospheric conditions. These tests led to the result that a limiting factor for the quality of the piston reconstruction is the instantaneous precision of the wind vector estimation, which is given by the spatial sampling of the WFS. Typically, the AO systems developed for infrared observations at \SI{8}{\meter}-class telescopes are equipped with sufficient WFS sampling for good wind estimation. The tests on LBT data were promising and mostly confirmed the results from the simulations, although the used data were not ideal to test the final interferometric performance. The piston reconstruction showed good results for around two thirds of the available data sets with good seeing conditions (r$_0$ $\leq$ \SI{30}{\centi\meter}). For these data, a reconstruction of the piston drift with an error smaller than \SI{3}{\micro\meter} was consistently achieved with our single-aperture verification method. As described, this result however does not fully describe the final interferometric performance, as the testing method itself is limited by wavefront reconstruction errors, which are not part of the P-REx piston drift estimator. We expect P-REx to perform significantly better in  a real interferometer. We used a virtual interferometer by masking the LBT entrance pupil down to two small ($\sim$ \SI{3.8}{\meter}) sub-apertures of the whole \SI{8.4}{\meter} aperture.  A full size \SI{8}{\meter} aperture WFS would have at least twice as many sub-apertures, thereby improving the wind vector estimation and the piston drift reconstruction. Despite these limitations, the results from the LBT data approved the assumptions on the structure of the turbulence and showed that under good atmospheric conditions, boiling is not a limiting factor for the performance of P-REx.

Based on the presented work, we conclude that the P-REx algorithm can limit the effective piston turbulence on second-timescales down below a few micrometers, which suggests benefits for interferometric observations at infrared science wavelength, and even for optical wavelength if coherencing is sufficient. This could be further supported by the fact that a two second timescale is probably not needed and a use over shorter timescale would ease the requirements on the system. As next step, we intend to further test P-REx with two-telescope interferometric data, in combination with fringe tracking measurements, and over a wider range of atmospheric conditions.

The final goal of our project is to implement the here developed techniques to an interferometric instrument. A good possibility for doing this would be the LBTI and the LINC-NIRVANA instruments at the LBT, as the binocular telescope operation with the very short baseline offers interferometric resolution without additional piston noise due to long delay lines and vibrating mirrors. Furthermore, LBT offers full real-time control of the vibrations in the telescope to its client instruments. Another possible instrument to apply P-REx to would be MATISSE, the mid-infrared spectro-interferometer for VLTI \citep{Lopez2014}. As MATISSE observes in the MIR, it imposes less constrains on the fringe stability and on the OPD noise. Increasing the coherence time for MATISSE operations would allow the observation of fainter targets, and at higher spectral resolution. Those experiments can be done without hardware upgrades, since both VLTI and LBT are already equipped with sufficiently performing AO systems. As discussed above, a further ideal application for P-REx would be an interferometer equipped with a laser-based GLAO system, similar to  LBT/ARGOS or VLT/AOF. A piston-predictive algorithm like P-REx in combination with such a laser-AO system is needed to eventually bring to direct infrared interferometry a similar increase of sky-coverage as was eventually achieved with laser-based AO wavefront control for single aperture science. This should be considered for the next generation of interferometric facilities, like the planet formation imager \citep[\href{http://www.planetformationimager.eu}{PFI}, see][]{Monnier2016}.

\section*{Acknowledgements}
We want to thank Simone Esposito and Alfio Puglisi for providing the LBT FLAO data used in in this work and the support in understanding and processing the datasets. The LBT is an international collaboration among institutions in the United States, Italy and Germany. LBT Corporation partners are: The University of Arizona on behalf of the Arizona Board of Regents; Istituto Nazionale di Astrofisica, Italy; LBT Beteiligungsgesellschaft, Germany, representing the Max-Planck Society, The Leibniz Institute for Astrophysics Potsdam, and Heidelberg University; The Ohio State University, and The Research Corporation, on behalf of The University of Notre Dame, University of Minnesota and University of Virginia.\\
This project has received funding from the European Union's Horizon 2020 research and innovation programme under grant agreement No 730890. This material reflects only the authors views and the Commission is not liable for any use that may be made of the information contained therein.

\bibliographystyle{mnras}
\bibliography{bibfile} 


\bsp	
\label{lastpage}
\end{document}